\documentclass[aps,prx,superscriptaddress,twocolumn]{revtex4-2}
\usepackage{graphicx,amsmath,amssymb,amsfonts,latexsym,color,dcolumn,bm}
\usepackage[T1]{fontenc}

\newcommand{\beq}{\begin{equation}}
\newcommand{\eeq}{\end{equation}}
\newcommand{\bea}{\begin{eqnarray}}
\newcommand{\eea}{\end{eqnarray}}
\providecommand{\abs}[1]{\left\lvert#1\right\rvert}

\providecommand{\bra}[1]{\langle #1 \rvert}
\providecommand{\ket}[1]{\lvert #1 \rangle}

\usepackage{amsfonts,amssymb}
\usepackage{dsfont}
\usepackage{physics}
\usepackage{tocvsec2}

\usepackage{appendix}

\usepackage{tikz}

\usepackage{amsmath}
\usepackage{bbold}
\usepackage{hyperref}
\hypersetup{
     colorlinks=true,      
    linkcolor=blue,        
    citecolor=blue,        
    filecolor=magenta, 
    urlcolor=black          
}

\usepackage{comment}

\begin{document}

\author{N. Fabre\footnote{nicolas.fabre@telecom-paris.fr}}
\affiliation{Telecom Paris, Institut Polytechnique de Paris, 19 Place Marguerite Perey, 91120 Palaiseau, France}
\begin{abstract}
We present a linear optical protocol for teleporting and correcting both temporal and frequency errors in two time-frequency qubit states. The first state is the frequency (or time-of-arrival) cat qubit, which is a single photon in a superposition of two frequencies (or time-of-arrival), while the second is the time-frequency Gottesman-Kitaev-Preskill (GKP) state, which is a single photon with a frequency comb structure. The proposed optical scheme could be valuable for reducing error rate in quantum communication protocols involving one of these qubits.
\end{abstract}

\title{Teleportation-based error correction protocol of  time-frequency qubits states}

\maketitle

\section{Introduction}
Quantum information can be encoded in various degrees of freedom of single photons, which can be described by either discrete or continuous variables (CV). Frequency (or energy) and time-of-arrival are natural pairs of conjugate quantum continuous variables in the single photon subspace, along with the transverse position and momentum degrees of freedom \cite{fabre_generation_2020, fabre:tel-03191301, tasca_continuous_2011, fabre_time_2022}. Discretizing the frequency or time-of-arrival into temporal or frequency bins, or performing the mode decomposition of the the continuous variable distribution of the single photon, can be experimentally motivated due to the finite resolution of detection devices or specific requirements of a quantum protocol, such as in quantum metrology for super-resolution \cite{PhysRevX.6.031033}. We should stress that in any dimensional single-photon encoding, photon losses do not correspond to a logical error. The second way of encoding information is through a particle-number sensitive encoding, which can be used to define physical systems with either discrete or continuous variables. In this encoding, CV corresponds to the quadratures of the electromagnetic field, {\it{i.e}}, the amplitude and phase of the quantum field, in a particular mode. With the particle-sensitive encoding, photon loss corresponds to a logical error. Mathematically, the quadrature of an electromagnetic field in a given mode can be treated as the continuous degree of freedom of a single photon \cite{fabre_time_2022}, as long as an auxiliary discrete mode is occupied by only one single photon. \\

Error correction code for continuous variables encoding is defined by discretizing them. Three bosonic qubit codes have been studied, such as the cat-code \cite{cochrane_macroscopically_1999, PhysRevX.9.041053,Albert_2019}, Gottesman, Kitaev and Preskill (GKP) code \cite{gottesman_encoding_2001,fluhmann_encoding_2019,PhysRevA.99.032344,campagne-ibarcq_quantum_2020,https://doi.org/10.48550/arxiv.2205.09781,PhysRevA.103.032409,8482307} and the binomial code \cite{michael_new_2016,hu_quantum_2019}. Cat and GKP codes are candidates for achieving universal quantum computation, see \cite{PhysRevX.9.041053,PRXQuantum.3.010329} and \cite{PhysRevLett.123.200502,bourassa_blueprint_2021}. GKP codes could be employed for building quantum repeaters \cite{rozpedek_quantum_2021}, and for sensing application \cite{Duivenvoorden_2017,Terhal_2016}. The mathematical analogy between time-frequency and quadrature CV allows defining time-frequency qubits states,  called time-frequency cat state \cite{fabre:tel-03191301, PhysRevA.102.023710} and time-frequency GKP state \cite{fabre_generation_2020,fabre:tel-03191301,https://doi.org/10.48550/arxiv.2301.03188}. Both of these codes are ways to discretize time-frequency continuous variables at the single photon level to define a qubit. CV or time-frequency CV codes possess an equivalent mathematical structure, they are common eigenvectors of non-commuting displacements operators \cite{fabre_generation_2020} and they are thus designed to be robust against small shift in one continuous variable (cat state) and the two canonically conjugated ones (GKP state). \\

In this paper, we start by reminding the mathematical structure of the time-frequency cat and GKP codes, and discuss the temporal and frequency errors from which they are designed to be robust against. We then analyse two different entanglement structures of time-frequency GKP state which can be generated experimentally, and can be interpreted as the entanglement of a noisy state of interest with a less noisy ancilla. However,  these entanglement structures which lead to a natural error correction strategy  for single photon encoding is difficult to realize experimentally with the current technology,  since it requires to perform a frequency entanglement operation between two single photons. Therefore, the standard method for error correction for quadrature qubits cannot be applied straightforwardly. Inspired by the teleportation-based error correction protocol for quadrature variables \cite{PhysRevA.102.062411,PhysRevResearch.3.033118}, we develop a teleportation-based error correction protocol of frequency qubits states, using only linear optical elements, and allows correcting both time and frequency variables at once. The error correction is naturally performed because the ancilla EPR state which assists to the teleportation is less noisy in the temporal and frequency domains compared to the state of interest.  Since the protocol requires the use of Bell's measurement, we also describe how to experimentally implement such a measurement for the two type of frequency qubits. The proposed protocol is intrinsically probabilistic but consist on the teleportation of non-orthogonal states \cite{sisodia_teleportation_2017}, instead of orthogonal ones \cite{PhysRevLett.70.1895, bouwmeester_high-fidelity_2000}. The non-orthogonality of the state reduces the efficiency of the teleportation protocol, and we mention that photon number resolving detectors can help increase the probability of success by reducing the number of rejected measurement events. The presented teleportation protocol is a new solution for correcting temporal broadening caused by dispersion effects affecting time-bin qubit states, thereby reducing the error rate of quantum communication protocols \cite{Zhong_2015,Jin:19,PhysRevApplied.14.014051}. \\

The paper is organized as follows. In Sec.~\ref{universal}, we provide a reminder of the definition of the time-frequency cat and GKP states and the reason of the experimental difficulty behind the natural error correction scheme of the time-frequency GKP states which requires frequency entanglement gates. In Sec.~\ref{concrete}, we explain the optical equivalent of the polarizing beam-splitter, a Mach-Zehnder interferometer, which allows separating spatially the two logical states of the time-frequency cat and GKP qubits.  Such an interferometer is crucial for implementing Bell's measurement in the time-frequency degree of freedom. In Sec.~\ref{sectionteleportation}, we present a teleportation-based error correction protocol for the time-frequency GKP state, which makes use of the Bell's measurement.  The protocol is probabilistic and can be achieved with current experimental devices. Finally, in Sec.~\ref{conclusion}, we summarize our results and present new perspectives.

\section{Time-frequency qubits states}\label{universal}

\subsection{Time-frequency cat state}
We will denote $\ket{\Omega}$ the vacuum state. A single photon state at frequency $\omega$ in the spatial port $a$ is denoted as $\ket{\omega}_{a}=\hat{a}^{\dagger}(\omega)\ket{\Omega}$. The frequency cat state as introduced in \cite{PhysRevA.102.023710}, is defined as the superposition of a single photon into two different frequency Gaussian distribution:
\begin{equation}\label{catstate}
\ket{\psi}=N_{\alpha\beta}(\alpha\ket{\omega_{1}}_{a}+\beta\ket{\omega_{2}}_{a})=N_{\alpha\beta}(\alpha \ket{0}_{a}+\beta \ket{1}_{a}),
\end{equation}
where $\ket{\omega_{1}}_{a}=\frac{1}{\sqrt{2\pi\sigma^{2}}} \int d\omega \text{exp}(-(\omega-\omega_{1})^{2}/2\sigma^{2}) \ket{\omega}_{a}$ and the normalization of the state is given by $1=N_{\alpha\beta}^{2}(\abs{\alpha}^{2}+\abs{\beta}^{2}+2\text{Re}(\alpha\beta^{*}) e^{-(\omega_{1}-\omega_{2})^{2}/2\sigma^{2}})$. This is a non-orthogonal qubit state as their overlap is ${}_a\bra{0}\ket{1}_{a}=\text{exp}(-(\omega_{1}-\omega_{2})^{2}/2\sigma^{2})$. 
Experimental proposal for manipulating such a state was proposed in \cite{lukens_frequency-encoded_2017,lu_controlled-not_2019}, with pulse shapers and electro-optic modulators were used in cascade. The results showed an operation fidelity close to 100\% but the successive optical elements decrease drastically the probability of single photon detection. The wavefunction of the time cat state is defined as:
\begin{equation}\label{timebinqubit}
\ket{\psi}=N_{\alpha\beta}(\alpha \ket{t_{1}}_{a}+\beta \ket{t_{2}}_{a}).
\end{equation}
Note that for avoiding to have a normalization constant depending in the coefficients $\alpha,\beta$, and writing the wavefunction in an orthogonal basis, we can employ the Gram-Schmidt decomposition procedure. The normalized orthogonal basis $\ket{a}$ and $\ket{b}$ can be written as:
\begin{equation}
\ket{a}=\ket{0}, \ \ket{b}=N[\ket{1}-\bra{0}\ket{1}\ket{0}],
\end{equation}
where $N=1/\sqrt{1-r^{2}}$ where $r=\abs{\bra{0}\ket{1}}$. The GKP input state Eq.~(\ref{GKPinput}) can be written in the orthogonal basis as:
\begin{equation}
\ket{\psi}=(\alpha+\beta\bra{0}\ket{1})\ket{a}+\frac{\beta}{N}\ket{b},
\end{equation}
where we have now $\abs{(\alpha+\beta\bra{0}\ket{1})}^{2}+\abs{\frac{\beta}{N}}^{2}=1$.\\

 The frequency entangled cat state, an EPR state could be written as:
\begin{align}\label{EPRfrequency}
\ket{\phi^{\pm}}=N_{\text{EPR}}(\ket{\omega_{1}\omega_{1}}_{ab}\pm\ket{\omega_{2}\omega_{2}}_{ab})\\
\ket{\psi^{\pm}}=N_{\text{EPR}}(\ket{\omega_{1}\omega_{2}}_{ab}\pm\ket{\omega_{2}\omega_{1}}_{ab})
\end{align}
where $N_{\text{EPR}}^{2}(2+2e^{-(\omega_{1}-\omega_{2})^{2}/2\sigma^{2}})=1$. When $\omega_{1}-\omega_{2} \gg \sigma$ we recover the normalization of an EPR state composed of orthogonal qubits $N_{\text{EPR}}=1/\sqrt{2}$. The frequency cat state can be produced by integrated optical waveguide \cite{https://doi.org/10.48550/arxiv.2207.10943}, and bulk system \cite{chen_hong-ou-mandel_2019}. The wavefunction of a temporal entangled EPR state Eq.~(\ref{EPRfrequency}) has the same mathematical structure, and such a quantum state can be produced by quantum dots  for instance \cite{jayakumar_time-bin_2014}. This type of quantum state has potential applications in quantum communications \cite{kim_quantum_2022}.

\subsection{Time-frequency GKP state}
\begin{figure*}
 \begin{center}
\includegraphics[width=0.8\textwidth]{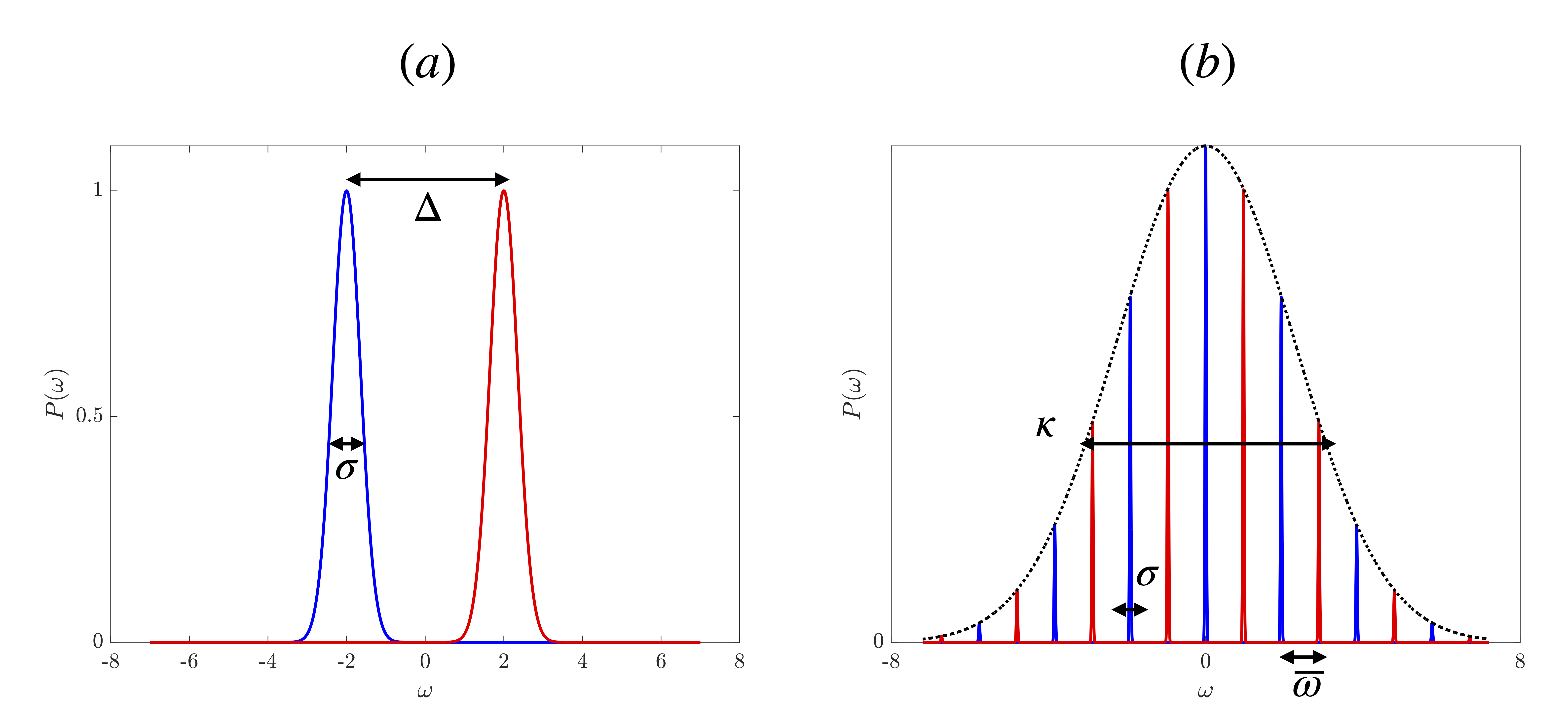}
\caption{\label{GKP}Probability distribution of the two time-frequency qubits. (a) Frequency cat state. $\sigma$ is the half-width of the Gaussian peak, and $\Delta$ is the spectral separation between the two logical states (b) Time-frequency GKP state. The width of each peak is $\sigma$, the envelope $\kappa$, and the periodicity of the comb state is $\overline{\omega}$. In the temporal domain, the periodicity of the state is $2\pi/\overline{\omega}$, the width of the envelope and the peak is $\sigma$ and $\kappa$ respectively. Frequency units are made dimensionless with respect to the spatial separation of the state (left) and the periodicity of the state (right).}
\end{center}
\end{figure*}

We define a frequency lattice of period $\overline{\omega}$. Centered on each of this interval, we define the ideal time-frequency GKP state as the following frequency comb at the single photon level:
\begin{equation}
\ket{\overline{\mu}_{\omega}}_{a}=\sum_{n\in\mathds{Z}} \ket{(2n+\mu)\overline{\omega}}_{a}
\end{equation}
where $\mu=0,1$ index the two logical states. Note that the equal weight superposition of the zero and the one logical time-frequency GKP states in the frequency domain are the zero and one in the temporal domain:
\begin{align}
\ket{\overline{+}_{\omega}}_{a}=\frac{1}{\sqrt{2}}(\ket{\overline{0}_{\omega}}_{a}+\ket{\overline{1}_{\omega}}_{a})=\ket{\overline{0}_{t}}_{a}\\
\ket{\overline{-}_{\omega}}_{a}=\frac{1}{\sqrt{2}}(\ket{\overline{0}_{\omega}}_{a}-\ket{\overline{1}_{\omega}}_{a})=\ket{\overline{1}_{t}}_{a}.
\end{align}
The periodicity of the state in the temporal domain: $\overline{\omega}=2\pi/\overline{\omega}$. Such a state is not physical since the state is an infinite sum of monochromatic state and will require an infinite energy to prepare it. The physical time-frequency GKP state can be built upon this ideal state by applying time and frequency noise, which are frequency and time displacement operations multiplied by Gaussian distribution which is detailed in \cite{fabre_generation_2020}. The wavefunction of the two logical states can be written as follows:
\begin{equation}
\ket{\mu_{\omega}}_{a}=N_{\mu}\sum_{n\in\mathds{Z}}\int d\omega G^{\kappa}(\omega)G^{\sigma}(\omega-(2n+\mu)\overline{\omega}) \ket{\omega}_{a}
\end{equation}
where $G$ are Gaussian functions representing the envelope of the comb of width $\kappa$ and the peaks of the comb of width $\sigma$. The frequency probability distribution of the grid state is represented in Fig.~\ref{GKP}. Alternatively for large comb $\overline{\omega}/\sigma \gg 1$ \cite{PhysRevA.94.022325}, we can write:
\begin{equation}
\ket{\mu_{\omega}}_{a}=N_{\mu}\sum_{n\in\mathds{Z}}c_{2n+\mu} \int d\omega G^{\sigma}(\omega-(2n+\mu)\overline{\omega}) \ket{\omega}_{a}
\end{equation}
with the envelope coefficients $c_{n}=\text{exp}(-(n\overline{\omega}/\kappa)^{2}/2)$. $N_{\mu}$ is the normalization constant found thanks to the relation $1=\abs{{}_a\bra{\mu_{\omega}}\ket{\mu_{\omega}}_{a}}^{2}=N_{\mu}^{2}\sum_{n\in\mathds{Z}} \abs{c_{2n+\mu}}^{2} \sqrt{\pi \sigma^{2}}$. \\
 
In general, the physical GKP state can be in a superposition of the two logical states:
\begin{equation}\label{GKPinput}
\ket{\psi}=N_{\alpha\beta}(\alpha\ket{0_{\omega}}_{a}+\beta\ket{1_{\omega}}_{a}),
\end{equation}
where $N_{\alpha\beta}=(\abs{\alpha}^{2}+\abs{\beta}^{2}+2\text{Re}(\alpha^{*}\beta \bra{0}\ket{1})))^{-1/2}$. The two logical states are not orthogonal when a Gaussian wavepacket enters in the frequency bin of its neighbour. The overlap ${}_a\bra{0_{\omega}}\ket{1_{\omega}}_{a}$ is different than zero and is equal to:
\begin{equation}
{}_a\bra{0_{\omega}}\ket{1_{\omega}}_{a}=e^{-\overline{\omega}^{2}/4\sigma^{2}} \frac{\sum_{n}c_{2n}c^{*}_{2n+1}}{(\sqrt{\sum_{n} \abs{c_{2n}}^{2}\sum_{n} \abs{c_{2n+1}}^{2})} }.
\end{equation}
 The full state is thus described by five importants parameters. The complex parameters $\alpha$ and $\beta$  where the quantum information is encoded, the frequency width $\sigma,\kappa$ and the periodicity of the state $\overline{\omega}$. If the state is not too much noisy, meaning that $ {}_a\bra{0_{\omega}}\ket{1_{\omega}}_{a}\sim 0$, then the normalisation condition of Eq.~(\ref{GKPinput}) is given by $\abs{\alpha}^{2}+\abs{\beta}^{2}=1$. Finally, in  \cite{fabre_generation_2020,fabre_time_2022}, we  define the time-of-arrival and frequency operators which do not commute and verify an Heisenberg algebra. This is mathematically equivalent to the non-commutativity of time-frequency displacement operators. This property leads to consider temporal and frequency bandwidth as quantum noise at the single photon level. \\

The GKP states which are defined as  the sum of squeezed states in a given mode  \cite{gottesman_encoding_2001,albert_performance_2018}, are designed to be robust against small shift in position and momentum, which can be caused by a Gaussian quantum channel,  but they are also robust against photon losses  \cite{albert_performance_2018}. On the other hand, time-frequency GKP states are designed to be robust against small shift in time and frequency. The major difference between GKP states and time-frequency GKP states is that photon losses do not result in errors for the latter. In general, temporal errors are the dominant source of errors, while frequency is considered a robust degree of freedom, as it is barely affected by linear physical processes. Additionally, GKP states can be used for fault-tolerant universal quantum computation \cite{bourassa_blueprint_2021}. Due to their mathematical similarity with time-frequency GKP states, it is expected that they would lead to the same mathematical result. However, the generation of non-Gaussian states using the degree of freedom of a single photon is relatively simple to implement experimentally. As a result, the experimental implementation of a time-frequency GKP state is considered straightforward. In contrast, creating entanglement gates between two single photons is a more challenging task. For particle-number sensitive encoding, non-Gaussian operations involving the quadrature degree of freedom can be difficult to implement, while two-mode Gaussian operations, such as with a beam splitter, are easier to perform.

 \subsection{Sources of time-frequency noise}

In this section, we discuss the physical processes that lead to temporal-spectral broadening or distorsion. Coherent and incoherent errors will lead to either pure and mixed state respectively.\\
  
Temporal errors for both codes arise from temporal spreading of each wavepacket composing the state due to linear dispersion effect, described by a coherent model (see for instance \cite{hong_dispersive_2018} for an example in the single photon regime). After such second order dispersion effect, the temporal width of the Gaussian wavepacket becomes $\tau=\tau_{0}\sqrt{1+\tau_{c}^{4}/\tau_{0}^{4}}$ where $\tau_{0}$ is the initial width of the pulse and $\tau_{c}=\sqrt{\beta_{2}L}$, $L$ being the length of the dispersive medium and $\beta_{2}$ the dispersion coefficient. Such a dispersion process is described by an unitary operation, and it can in principle be undone by a reverse transformation. However, it requires the knowledge of the full characterization of the propagation channel. For the time-frequency GKP state, the dispersive effect not only lead to a temporal spreading of each wavepacket, but also lead to the formation of replica temporal images, called the temporal Talbot effect (see for instance  \cite{maram_spectral_2013}). One has to consider specific length of the fiber or dispersive coefficient to recover the initial state, which will be also temporally broadened. Polarizing mode dispersion is one of incoherent temporal broadening \cite{antonelli_pulse_2005,poon_polarization_2008,gordon_pmd_2000}. Due to a coupling between the polarization and frequency degree of freedom, if the polarization is not measured, the single photon state becomes mixed in frequency \cite{chang-hua_polarization_nodate}. Thus, the error correction protocol that will be presented in Sec.~\ref{errorentangle} becomes particularly relevant because we can not cancel the error simply by a unitary operation.  \\

Frequency noise that causes spectral broadening while the spectral distribution remains Gaussian, is not typically dominant at the single-photon level. Spectral broadening induced by the self-phase modulation effect  \cite{matsuda_deterministic_2016} results from the accumulated phase $\phi_{NL}(t) = \frac{2\pi}{\lambda} n_{2} I(t)L$, where $n_{2}$ is the non-linear refractive index, $L$ is the length of the medium, and depends on the intensity $I(t)$ of the field, leading to a non-Gaussian spectral distribution. At the single-photon level, this non-linear process does not occur naturally. In \cite{fabre_generation_2020}, we also argue that frequency noise arises from frequency broadening caused by the generation of photon pairs itself, and we describe one method to correct such a noise. There are numerous processes that can distort the spectral distribution of single photons, such as distortions caused by frequency shifts induced by electro-optic modulators \cite{Kurzyna_2022,PhysRevLett.129.123605}, or the presence of a filter during single photon heralding.\\

An error correction protocol is implemented to mitigate the effects of potential errors from various sources. Its objective is to restore the Gaussian distribution of each frequency peak. This is because Gaussian probability distributions are better understood for setting confidence intervals and error thresholds, as discussed in \cite{fukui_high-threshold_2018}. Both qubits are sensitive to temporal and frequency broadening. While the time-frequency cat state is designed to be robust against errors over one variable, the time-frequency GKP state can correct errors along both orthogonal variables. It is not necessary to use the time-frequency GKP state if the main error is in the temporal domain. We now develop two error correction methods, one based on a direct frequency entanglement between the noisy state of interest and a less noisy ancilla Sec.~\ref{errorentangle}, the other method by using only linear optics and also less noisy ancilla Sec.~\ref{sectionteleportation}.

\subsection{Time-frequency entangled GKP state and error correction protocols}\label{errorentangle}
 Error correction of continuous variables states can be done with a Steane error correction protocol, for the quadrature degree of freedom \cite{seshadreesan_coherent_2021} and for the time-frequency one \cite{fabre_generation_2020}. For the two encodings, the protocol consists of entangling the state of interest with one less noisy ancilla (a $\ket{+}$ logical state in one variable, time or frequency), with a beam splitter (resp. with frequency beam-splitter that will be explicated below), and performing one homodyne (resp. single photon frequency measurement) detection at the spatial output of the ancilla for correcting the error along one variable. The protocol is repeated to correct errors in the orthogonally (or canonically) conjugated variable. To achieve this, one must entangle the state of interest with a less noisy state using a $\ket{+}$ state in the canonically conjugated variable compared to the first step. Then, perform the entangling operation, project a measurement in the canonically conjugated variable (compared to step one), and finally conduct a conditional displacement operation. Important tools for quantifying the threshold of noise from which it is still possible to correct the GKP states, using such a Steane error correction protocol was done in  \cite{fukui_high-threshold_2018}. Figures of merit for quantifying the probability of measuring the one logical state while this is the zero  that should be obtained  was done in \cite{PhysRevA.73.012325,fukui_high-threshold_2018}. We will refer to these figures of merit, when we will note one logical state is more (or less) noisy than the other.  From now on, we develop two entanglement structures of time-frequency GKP state that can be generated in the laboratory, and how to perform the error correction in each case. \\

\begin{figure*}
 \begin{center}
\includegraphics[width=1\textwidth]{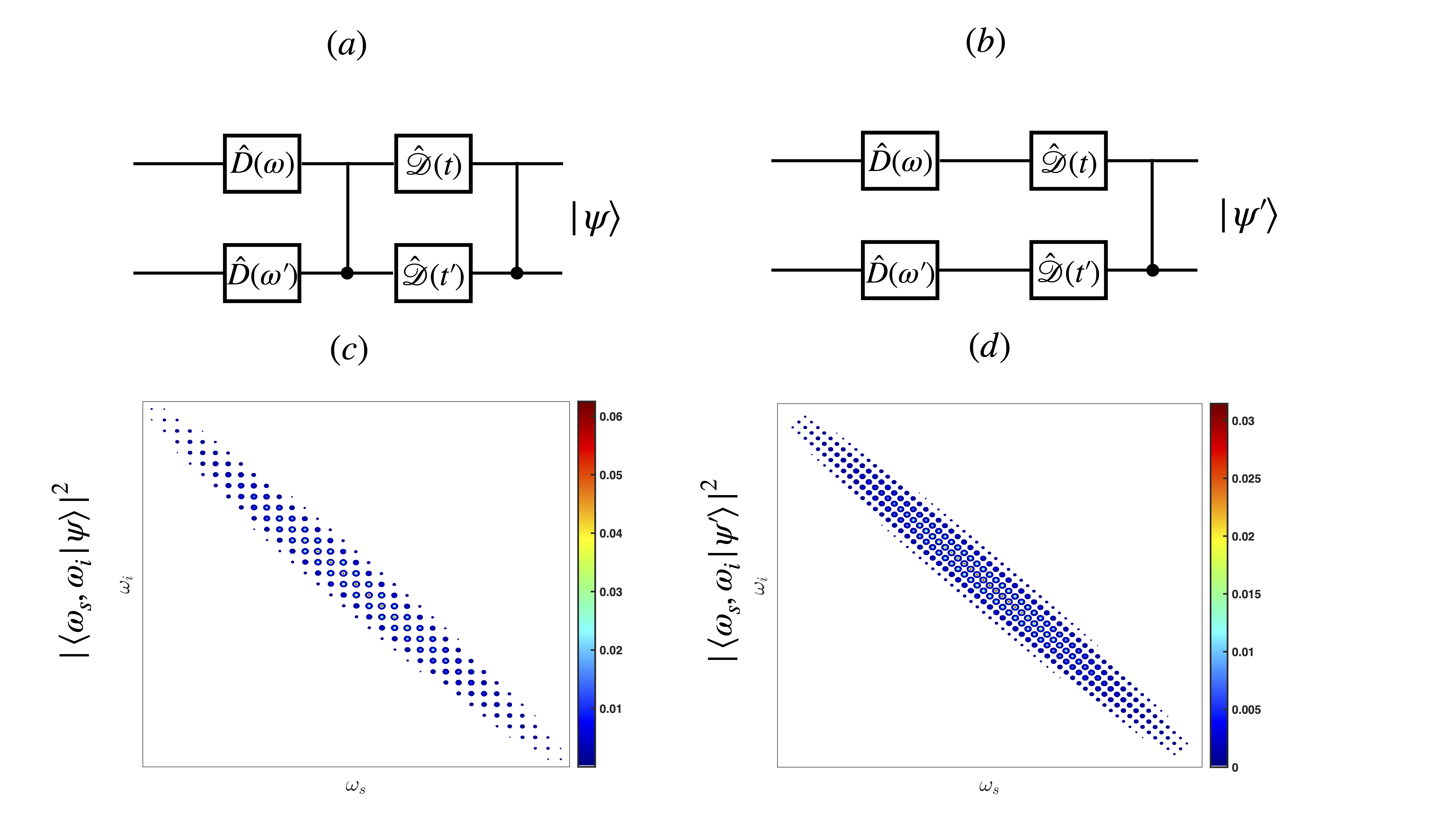}
\caption{\label{cestceci}Different joint spectral intensity $\abs{\bra{\omega_{s},\omega_{i}}\ket{\psi}}^{2}$ of time-frequency entangled GKP states (c),(d) that can be generated experimentally and their quantum circuit representation (a), (b). The situation (a) corresponds to a photon pair produced by a SPDC process, while (b) corresponds to two single photons with a frequency comb structure produced by two independent processes which are then frequency entangled. The position of frequency beam-splitter operations modifies the periodicity of the grid state from a factor $\sqrt{2}$. $\hat{D}(\omega)$ and ${\cal{\hat{D}}}(t)$ are frequency and temporal displacement operations defined in \cite{fabre_generation_2020}. }
\end{center}
\end{figure*}

The first entanglement structure of time-frequency GKP state that we can studied is the one obtained by using a spontaneous parametric down conversion process (SPDC) from a non-linear crystal placed into an optical cavity  \cite{fabre_generation_2020,maltese_generation_2020}. The corresponding wavefunction can be cast as:
\begin{equation}\label{firstGKP}
\ket{\psi}=\iint d\omega_{s} d\omega_{i} f_{+}(\omega_{+})f_{-}(\omega_{-})f(\omega_{s})f(\omega_{i}) \ket{\omega_{s},\omega_{i}}.
\end{equation}
The functions $f_{\pm}$ model respectively the energy conservation and the phase-matching of the SPDC process, and $f$ models the cavity function. The joint spectrum intensity is represented in Fig.~\ref{cestceci}(c). Instead of the traditional view where the wavefunction of two photons is written by applying a quadratic Hamiltonian on the vacuum state, it can also be expressed with a quantum circuit representation as shown in Fig.~\ref{cestceci}(a). This starts with two ideal separable GKP states (fictitious), which undergo initial frequency broadening, which can be interpreted as a frequency noise \cite{fabre_generation_2020}. Then, the state is entangled through the gate performed by the non-linear crystal:
\begin{equation}\label{FBS}
\hat{U}\ket{\omega_{s},\omega_{i}}=\ket{\frac{\omega_{s}+\omega_{i}}{\sqrt{2}},\frac{\omega_{s}-\omega_{i}}{\sqrt{2}}}
\end{equation}
that we called a frequency beam-splitter by analogy with the beam-splitter which acts mathematically similarly into the quadrature position-momentum degree of freedom  \cite{doi:10.1080/09500340.2022.2073613,fabre_time_2022}. The state then undergoes a temporal broadening which can be interpreted as a temporal noise, and a final frequency beam-splitter operation is performed. The mathematical reason behind these four successive operations is that the envelop of the grid state are function of the collective variables $\omega_{\pm}$, while the cavity of the local variable $\omega_{s,i}$.  The form of the frequency entanglement between the two single photon grid state generated by the non-linear process (see Eq.~(\ref{FBS})) allows for the reduction of the temporal broadening of one of the single photons by performing a temporally-resolved measurement of the other. The form of the pure wavefunction after the conditioned operation has been written in \cite{fabre_generation_2020}.
After correcting the single photon state in the temporal domain, the next step is to correct the state in the frequency domain. For that, we must first entangle the single photon to correct with a less noisy ancilla single photon state using the entanglement gate Eq.~(\ref{FBS}). Since now we have two separable single photon states, we can not directly entangle them by using a non-linear crystal, which is non-efficient. Nevertheless, such a frequency entangled gate could be implemented with a quantum emitter embedded into a waveguide, which assists to the interaction between the two single photons \cite{le_jeannic_dynamical_2022,PhysRevLett.126.023603}. \\

The second entanglement structure of time-frequency GKP state that can be considered, is to start with two initially separable GKP states, with one being less noisy than the other. These states are then entangled using Eq.~(\ref{FBS}). The resulting wavefunction is:
\begin{equation}\label{secondGKP}
\ket{\psi'}=\iint  d\omega_{s} d\omega_{i} f_{+}(\omega_{+})f_{-}(\omega_{-})f(\omega_{+})f(\omega_{-}) \ket{\omega_{s},\omega_{i}}.
\end{equation}
In this new spectral function (see Fig.~\ref{cestceci}(d)), the cavity function is now dependent on the collective variables, thus the periodicity of the grid state is not the same as in Eq.~(\ref{firstGKP}). The corresponding quantum circuit representation is pictured in  Fig.~\ref{cestceci}(b).  The error correction protocol is followed by a temporally-resolved measurement followed by a conditional displacement operation to correct only the temporal noise in the state of interest. A second entanglement operation is performed using a less noisy ancilla state, followed by a resolved-frequency measurement followed by a conditional displacement operation to correct the frequency noise. The difference in the joint spectral amplitude of the photon pairs results - Eq.~(\ref{firstGKP}) and Eq.~(\ref{secondGKP}) -  in a different wave function when one of the photons undergoes a temporally-resolved (and frequency) measurement, and a conditional displacement operation. \\

\section{Spatial separation of the two logical time-frequency qubit states}\label{concrete}
In this section, we develop the equivalent of the polarizing beam-slitter operation for the time-frequency cat and GKP state, which is the crucial optical component for the teleportation-based error correction described in the next section. Such an optical element will be called the frequency qubit beam-splitter (FQBS) in what follows.

\subsection{Spatial separation of the two logical time-frequency cat states}\label{spatialtwocolors}
In this section, we explicit how to separate spatially the two Gaussian wavepackets with linear optics, using a Mach-Zehnder interferometer. The frequency cat state as described by Eq.~(\ref{catstate}) is introduced into a balanced beam-splitter and the associated wavefunction is:
\begin{equation}
\ket{\psi}=\frac{1}{2}(\ket{\omega_{1}}_{a}+\ket{\omega_{2}}_{a}+\ket{\omega_{1}}_{b}+\ket{\omega_{2}}_{b}).
\end{equation}
We assumed for simplicity that the two frequency state are well separated $\omega_{1}-\omega_{2} \gg \sigma$.
Then, a pulse shaper is placed at the spatial port $b$, described by the following unitary operation:
\begin{equation}
\hat{U}\ket{\omega_{1}}_{b}=\ket{\omega_{1}}_{b}, \ \hat{U}\ket{\omega_{2}}_{b}=e^{i\phi} \ket{\omega_{2}}_{b}.
\end{equation}
Such an operation can be implemented for instance by mapping the spectral to the spatial degree of freedom with a grating, a spatial light modulator at the focal length of two lenses, and then performing back the mapping from the spatial to spectral degree of freedom \cite{fabre:tel-03191301,PhysRevA.94.063842}. We assume that the frequency peaks are spaced enough so that the pulse shaper acts on each logical state independently. The two spatial paths are then recombined to another balanced beam-splitter. The output state has the final form:
{\small{\begin{equation}
\ket{\psi}=\frac{1}{\sqrt{2}} \ket{\omega_{1}}_{a}\ket{\Omega}_{b}+\frac{1}{2\sqrt{2}}((1+e^{i\phi}) \ket{\omega_{2}}_{a}\ket{\Omega}_{b}+(1-e^{i\phi}) \ket{\Omega}_{a}\ket{\omega_{2}}_{b}).
\end{equation}}}
If $\phi=\pi$, the two logical states are spatially separated $\ket{\psi}=\frac{1}{\sqrt{2}}(\ket{\omega_{1}}_{a}\ket{\Omega}_{b}+\ket{\Omega}_{a}\ket{\omega_{2}}_{b})$. This optical interferometer is the equivalent of the polarizing beam-splitter that separates the vertical and horizontal polarization of optical fields into two distinct spatial paths.

\subsection{Spatial separation of the two logical time-frequency GKP states}\label{separationGKP}

In this part, we introduce how to separate spatially the odd and the even frequencies with a Mach-Zehnder interferometer. Such a scheme was already proposed for manipulating large quadrature position-momentum continuous variables cluster states \citep{PhysRevA.94.032327}, and was described in \cite{fabre:tel-03191301,https://doi.org/10.48550/arxiv.2301.03188}.\\

\begin{figure*}
 \begin{center}
\includegraphics[width=1\textwidth]{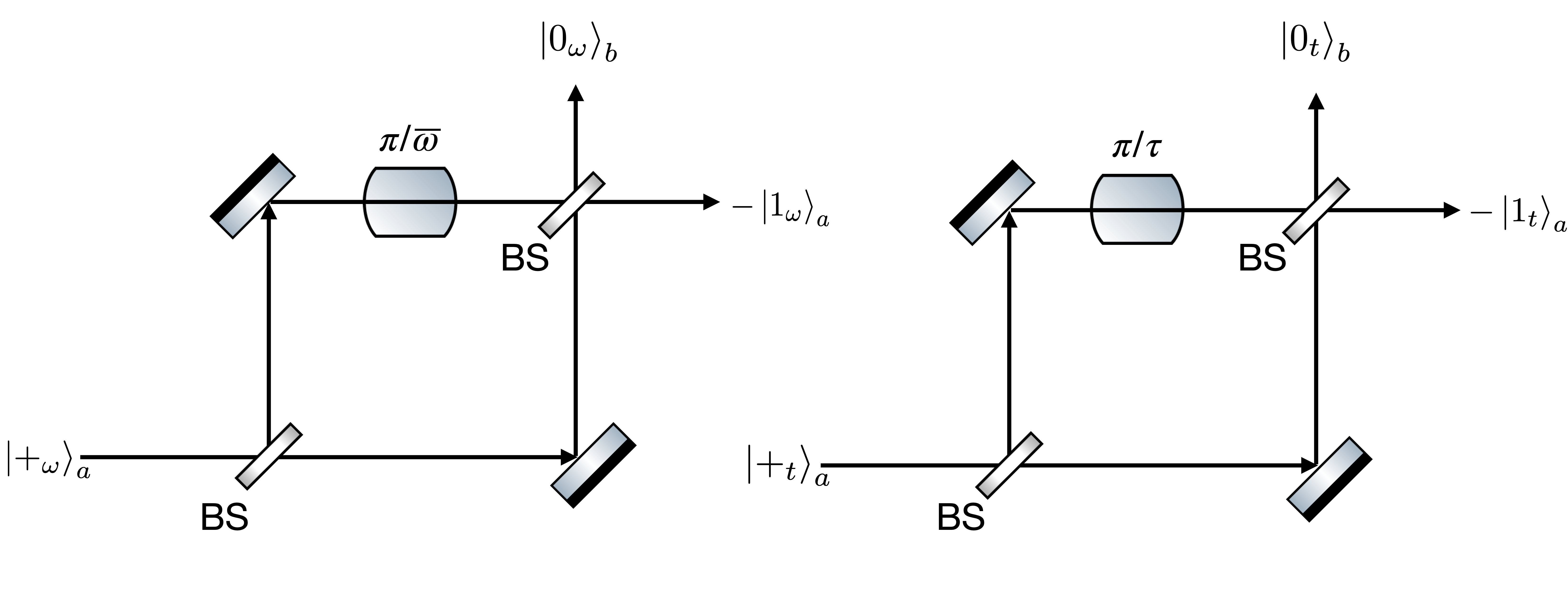}
\caption{\label{scheme}Spatial separation of the odd and even peaks of the time-frequency GKP state with a Mach-Zehnder interferometer in the frequency (left) and in the temporal (right) domain. BS stands for balanced beam-splitter. The spatial separation is imperfect when the states are no longer perfectly monochromatic.}
\end{center}
\end{figure*}

We start with a time-frequency GKP state with a finite envelop but with infinitely narrow frequency width $\ket{\tilde{+}}=\frac{1}{\sqrt{2}}(\ket{\tilde{0}}_{a}+\ket{\tilde{1}}_{a}$), and is introduced into a balanced beam-splitter. The spatial output port of the beam-splitter is noted $a$ and $b$. A time-shift operation is performed in spatial port $b$, and then the two spatial ports are recombined into a balanced beam-splitter (see Fig.~\ref{scheme}). The final wave function can be written as:
{\small{\begin{equation}
\ket{\psi}=\frac{1}{2}\sum_{n\in\mathds{Z}}[c_{n}(e^{in\overline{\omega} t}-1)\ket{n\overline{\omega}}_{a}\ket{\Omega}_{b}-(e^{in\overline{\omega} t}+1)\ket{\Omega}_{a}\ket{n\overline{\omega}}_{b}].
\end{equation}}}
If we set  $t=\pi/\overline{\omega}$, after the second beam-splitter the wavefunction is (see \cite{fabre:tel-03191301}):
\begin{equation}
\ket{\psi}=\frac{1}{\sqrt{2}}(-\ket{1}_{a}\ket{\Omega}_{b}+\ket{\Omega}_{a}\ket{0}_{b}).
\end{equation}
The odd and even frequency components are spatially separated, allowing for individual manipulation, such as correcting the phase accumulation of the one logical state. It is possible to achieve the same result in the temporal domain by shifting the frequency instead of the time, as shown in Fig.~\ref{scheme}.
 
\begin{figure*}
 \begin{center}
\includegraphics[width=1\textwidth]{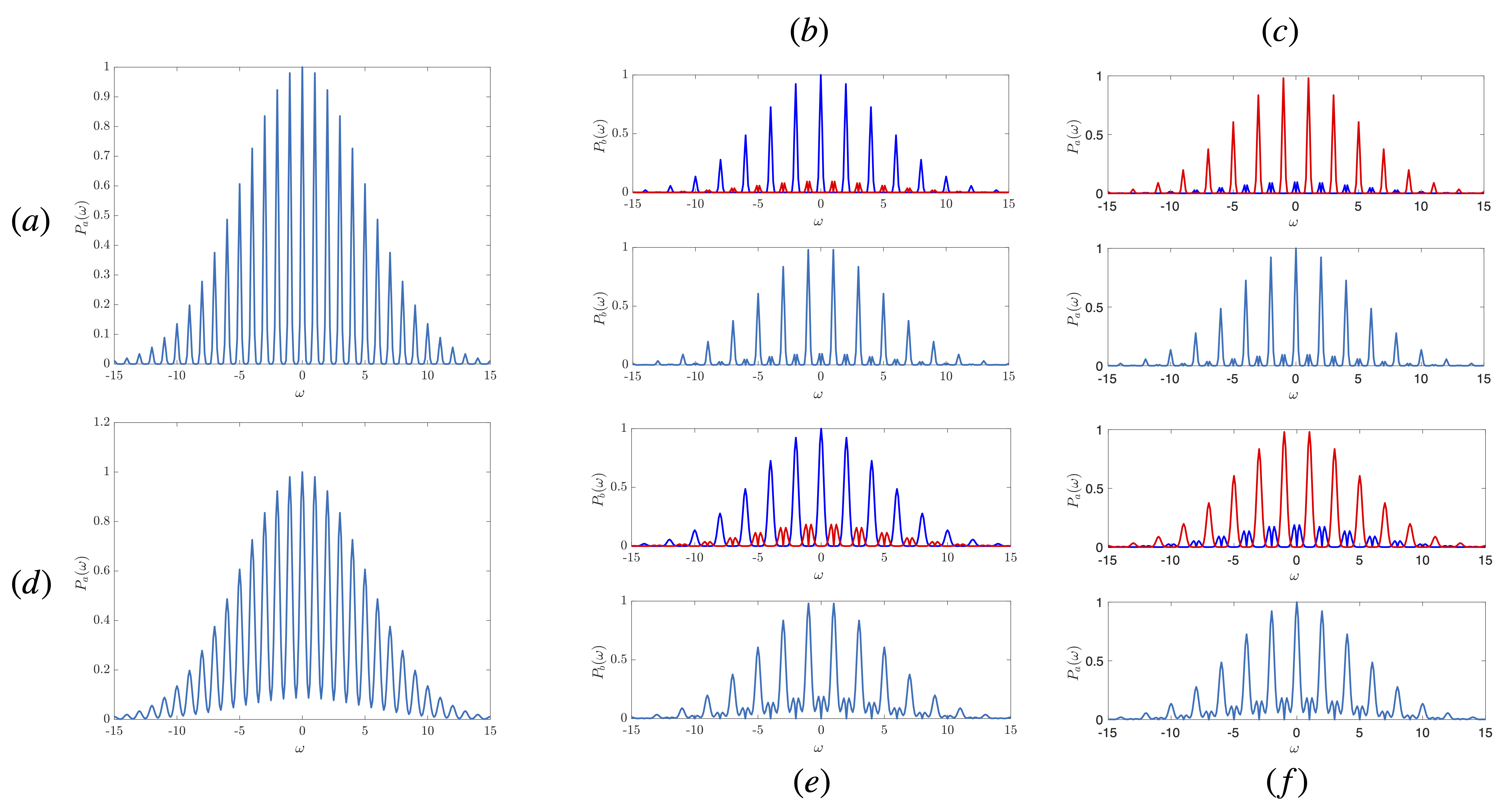}
\caption{\label{PFBS11}(a) Input state of the Mach-Zehnder interferometer  for $0.1\overline{\omega}=\sigma$ and $\kappa=0.1$. (b), (c) Probability distribution in the spatial port $b$ (resp. (a)) if we start with either the zero (blue) or one (one) logical state or with the equal superposition of the zero and one logical state. Their overlap is not important enough to observe an interference effect. (d) Input state of the Mach-Zehnder interferometer for $0.2\overline{\omega}=\sigma$ and $\kappa=0.1$. (e), (f) Probability distribution in the spatial port $b$ (resp. (a)) if we start with either the zero (blue) or one (one) logical state or with the equal superposition of the zero and one logical state. Their overlap is important enough to observe an interference effect to see a distorsion of the state which leaks to the incorrect spatial port.}
\end{center}
\end{figure*}

If each frequency peak is not infinitely narrow, then the output wavefunction after the Mach-Zehnder interferometer can be written as:
\begin{multline}\label{imperfectpfbs}
\ket{\psi}=\frac{1}{2}\sum_{n\in\mathds{Z}}c_{n}[(\int d\omega (e^{i\pi\omega/\overline{\omega} }-1)  G^{\sigma}(\omega-n\overline{\omega}) \ket{\omega}_{a}\ket{\Omega}_{b}\\
-(\int d\omega (e^{i\pi\omega/\overline{\omega} }+1)  G^{\sigma}(\omega-n\overline{\omega})\ket{\Omega}_{a} \ket{\omega}_{b}.
\end{multline}
The corresponding probability frequency distribution at spatial port $a$ and $b$ are
\begin{align}
P_{a}(\omega)=\frac{1}{4}\abs{\sum_{n\in\mathds{Z}}c_{n}(e^{i\pi\omega/\overline{\omega} }-1)  G^{\sigma}(\omega-n\overline{\omega})}^{2},\\
 P_{b}(\omega)=\frac{1}{4}\abs{\sum_{n\in\mathds{Z}}c_{n}(e^{i\pi\omega/\overline{\omega} }+1)  G^{\sigma} (\omega-n\overline{\omega})}^{2}.
 \end{align}
 We represent in Fig.~\ref{PFBS11}(a) (b) the spectral distribution of two $\ket{+}$ states for $\sigma=0.1\overline{\omega}$ and $\sigma=0.2\overline{\omega}$ along with the output probability distribution after the Mach-Zehnder interferometer in Fig.~\ref{PFBS11}(c),(d). We observe that when $\sigma = 0.1\overline{\omega}$, it is valid to consider the zero and one logical states independently since they do not interfere due to their central frequencies being too far apart for overlap. However, the spatial separation is imperfect, the one logical state does not emerge from the correct spatial port. When $\sigma=0.2\overline{\omega}$ (see Fig.~\ref{PFBS11}), there is an interference term between the zero and one logical state because they now overlap significantly. The resulting state is outside the GKP subspace, it can be seen with the distorsion and because the probability at the center of the odd and even frequency bins is zero. The addition of a periodic frequency filter can enhance the projection into the GKP subspace and the choice of a frequency width would be crucial for rejecting the logical states emerging from the incorrect spatial port.\\

In Appendix \ref{Twophoton}, we also investigate the separation of the odd and even components of the comb when the two-photon state is an input of the frequency qubit beam-splitter. This is relevant to the teleportation-based error correction protocol because the state to be teleported is combined with a single photon from an EPR pair during the Bell measurement. The spatial separation is, in that case, also completely effective when each Gaussian distribution approaches a Dirac distribution, and not otherwise. Explicitly, for the two photons states we obtain: $\ket{0\tilde{1}}_{aa}\rightarrow \ket{0_{G}}_{a}\ket{\tilde{1}_{G}}_{a'}+\ket{0_{E}}_{a'}\ket{\tilde{1}_{E}}_{a}$, whose the corresponding expressions in given in Appendix \ref{Twophoton}. The case to combine two different single photon states also plays a role in quantum communication scenarios, where the second qubit is from an attacker which tries to collect the information about the qubit carrying the information of interest.  \\

\section{Teleportation-based error correction of time-frequency qubits states}\label{sectionteleportation}
In this section, we propose a protocol for correcting and teleporting frequency qubit states without relying on frequency entangling operations. This protocol is similar to the one used for teleporting polarization qubit states as described in \cite{lutkenhaus_bell_1999}, and is inspired from the GKP analog \cite{PhysRevA.102.062411,PhysRevResearch.3.033118}. Since the EPR state has a lower level of noise in both temporal and frequency variables, the protocol includes an additional component for correcting errors in the state being teleported.\\

The single photon state $\ket{\psi}$ to be teleported and corrected is described by the wavefunction $\ket{\psi}=N_{\alpha\beta}(\alpha \ket{0}_{a}+\beta\ket{1}_{a})$, where the two logical state are either the time-frequency cat or GKP qubits (see Eq.~(\ref{GKPinput})).  The wavefunction of the entangled EPR time-frequency state in spatial port $b$ and $c$ which assists for the teleportation and the correction is:
\begin{equation}
\ket{\phi^{+}}_{bc}=N_{\text{EPR}}(\ket{\tilde{0}\tilde{1}}_{bc}+\ket{\tilde{1}\tilde{0}}_{bc})
\end{equation}
composed of logical states which are less noisy, indicated with the tilde notation, than the state to be teleported. Upon completion of the protocol, the single-photon state $\ket{\psi}$ will be localized to the spatial port $c$, and the correction is automatically performed since the EPR state is less noisy than the state of interest.\\

\begin{figure*}
 \begin{center}
\includegraphics[width=1\textwidth]{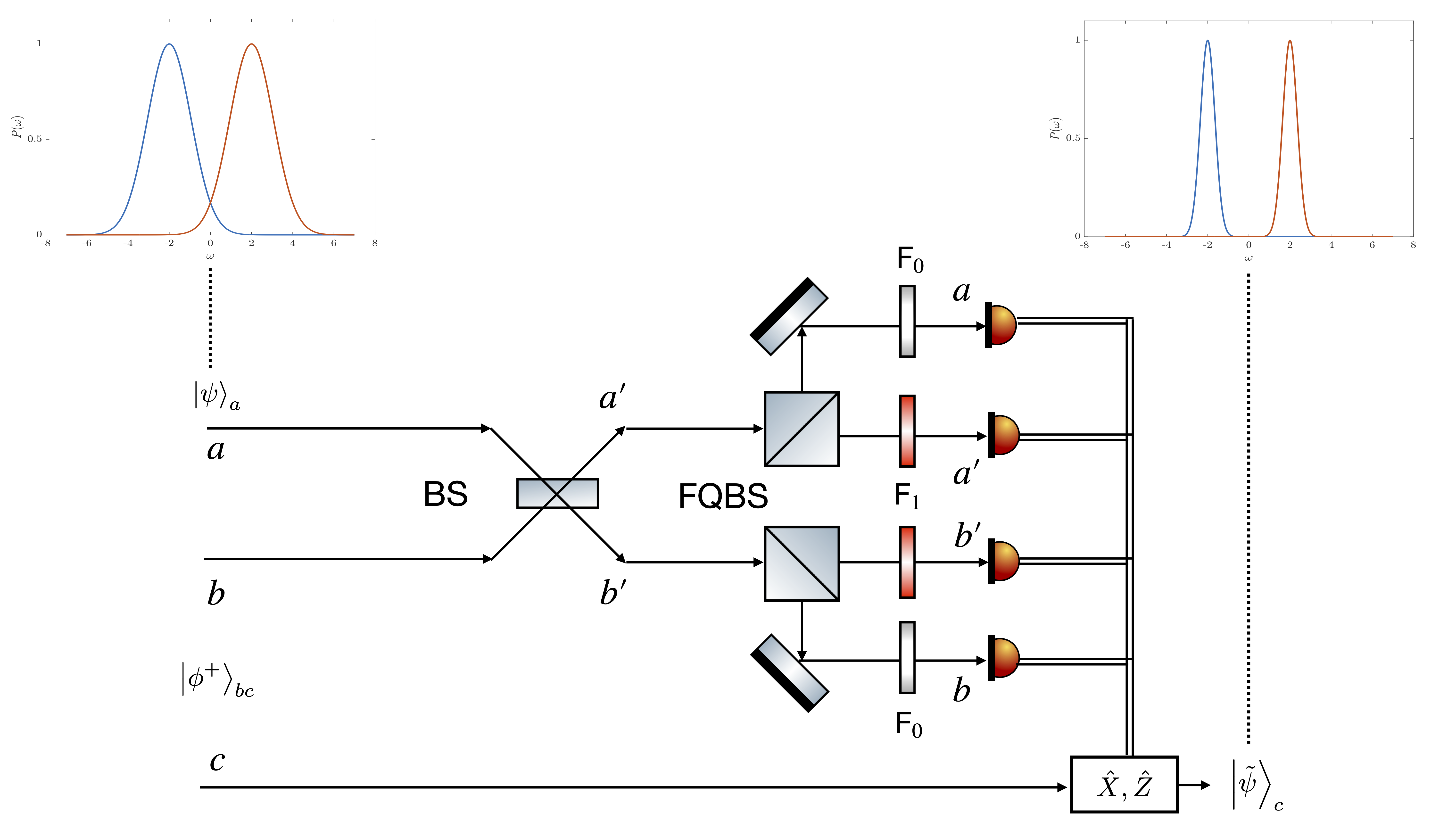}
\caption{\label{teleportationscheme} Schematic of the teleportation-based error correction of frequency qubit states. The state to be teleported and corrected is located in spatial port A. An EPR frequency qubit states which is less noisy than the state of interest, is at spatial ports B and C. FQBS stands for frequency qubit beam-splitter. $F_{0,1}$ are frequency filters with central frequency matching either the zero and one logical state, and with a frequency width equal to that of the EPR state. Depending on which detectors have measured coincidences, Pauli operations $\hat{X},\hat{Z}$ must be performed to recover the state of interest. The frequency qubit states before and after the teleportation are represented. The zero logical (resp. one) state has a blue (resp. red) color.}
\end{center}
\end{figure*}

 We now write the protocol for the error correction and teleportation of the time-frequency cat state, considering that the Bell's measurement perfectly separate the two logical states.  This protocol allows correcting frequency errors affecting a frequency qubit states. The protocol is represented in Fig. \ref{teleportationscheme}, and we will now proceed to write the evolution of the wavefunction at each step of the protocol. The initial wavefunction, composed of the state to be corrected and teleported, and the EPR state, written in the Bell's basis is:
\begin{multline}
\ket{\psi}=\frac{N_{\alpha\beta}N_{\text{EPR}}}{2}(\ket{\omega_{1}\tilde{\omega}_{1}}+\ket{\omega_{2}\tilde{\omega}_{2}})(\alpha \ket{\tilde{\omega_{1}}}_{c}+\beta \ket{\tilde{\omega_{2}}}_{c})\\
+(\ket{\omega_{1}\tilde{\omega}_{1}}-\ket{\omega_{2}\tilde{\omega}_{2}})(\alpha \ket{\tilde{\omega_{1}}}_{c}-\beta \ket{\tilde{\omega_{2}}}_{c})\\
+(\ket{\omega_{1}\tilde{\omega}_{2}}+\ket{\omega_{2}\tilde{\omega}_{1}})(\beta \ket{\tilde{\omega_{1}}}_{c}+\alpha \ket{\tilde{\omega_{2}}}_{c})\\
+(-\ket{\omega_{1}\tilde{\omega}_{2}}+\ket{\omega_{2}\tilde{\omega}_{1}})(\beta \ket{\tilde{\omega_{1}}}_{c}-\alpha \ket{\tilde{\omega_{2}}}_{c}).
\end{multline}
The single photon state and one member of the EPR pair are then combined into a beam-splitter, followed by two parity-frequency beam-splitter are placed in spatial port $a'$ and $b'$. The first and second Bell's state are transformed as:
\begin{multline}
\frac{N_{\text{EPR}}}{2}[\ket{\omega_{1}\tilde{\omega}_{1}}_{a}-\ket{\omega_{1}}_{a}\ket{\tilde{\omega}_{1}}_{b}+\ket{\omega_{1}}_{b}\ket{\tilde{\omega}_{1}}_{a}-\ket{\omega_{1}\tilde{\omega}_{1}}_{b}\\
+\ket{\omega_{2}\tilde{\omega}_{2}}_{a}-\ket{\omega_{2}}_{a}\ket{\tilde{\omega}_{2}}_{b}+\ket{\omega_{2}}_{b}\ket{\tilde{\omega}_{2}}_{a}-\ket{\omega_{2}\tilde{\omega}_{2}}_{b}].
\end{multline}
The presence of a single photon in each port is a consequence of the distinguishability of the photons. While if the photons were indistinguishable, when the two logical states are orthogonal as it is the case for the polarization encoding, only bunching event will be measured. In order to suppress this coincidence events that leads to errors in the teleportation protocol, the utilisation of frequency filters with the same frequency width (envelope and peak) as the EPR state is a potential solution. Explicitly, after the filtering operation, the state becomes $\ket{\tilde{\omega}_{1}\tilde{\omega}_{1}}_{a}-\ket{\omega_{1}\tilde{\omega}_{1}}_{b}+\ket{\tilde{\omega}_{2}\tilde{\omega}_{2}}_{a}-\ket{\omega_{2}\tilde{\omega}_{2}}_{b}$, which leads only to two bunching events, that are ignored if single photon detectors are used. The same analysis can be employed for the second Bell's state.\\

For the third and fourth Bell's state, they are transformed as:
\begin{multline}
\frac{N_{\text{EPR}}}{2}[ \pm \ket{\omega_{1}\tilde{\omega}_{2}}_{aa'}\mp\ket{\omega_{1}\tilde{\omega}_{2}}_{ab'}\pm\ket{\omega_{1}\tilde{\omega}_{2}}_{ba'}\mp \ket{\omega_{1}\tilde{\omega}_{2}}_{bb'}\\
+\ket{\omega_{2}\tilde{\omega}_{1}}_{a'a}-\ket{\omega_{2}\tilde{\omega}_{1}}_{a'b}+\ket{\omega_{2}\tilde{\omega}_{1}}_{b'a}-\ket{\omega_{2}\tilde{\omega}_{1}}_{b'b}].
\end{multline}
The use of frequency filter is again imperative, since the measurement of coincidence once filtered of $a,a'$ (or $b,b'$) permits to ensure that the quantum state $N_{\alpha\beta}(\beta \ket{\tilde{\omega_{1}}}+\alpha \ket{\tilde{\omega_{2}}})$ has been teleported. In the same way, the measurement of the coincidence of $a,b'$ (or $b,a'$) allows to teleport the state $N_{\alpha\beta}(\beta \ket{\tilde{\omega_{1}}}-\alpha \ket{\tilde{\omega_{2}}})$. The receiver sends to spatial port $c$ which detectors have measured coincidences, and then a product of Pauli matrix gates must be applied to recover the initial state of interest. Pauli matrices for the time-frequency cat states are frequency and temporal shifts operations \cite{fabre_generation_2020}, which can be implemented by either a electro-optical modulator \cite{Kurzyna_2022,PhysRevLett.129.123605} and a delay line respectively. The full optical scheme of the teleportation-based error correction protocol is represented in Fig.~\ref{teleportationscheme}, along with a illustration of the effect of the error correction for qubit cat states.\\

The corresponding probability of each event is $P=\frac{1}{8(1+e^{-\Delta^{2}/2\sigma^{2}})}$, and the overall probability success of the teleportation is
\begin{equation}
P=\frac{1}{2(1+e^{-\Delta^{2}/2\sigma^{2}})},
\end{equation}
which is then lower than 50 \%, since only linear optics is used \cite{lutkenhaus_bell_1999}, and because the non-orthogonality of the encoding further decrease the probability of success. Note that the use of photon-number-resolving (PNR) detectors will able to not discard the bunching events coming from the first and second Bell's state. With the use of such a PNR detector, the overall probability of success of the teleportation protocol is:
\begin{equation}
P_{\text{PNR}}=\frac{3}{4(1+e^{-\Delta^{2}/2\sigma^{2}})}
\end{equation}
and thus allows to increase the probability of success of the protocol despite the non-orthogonality of the state. Experimentally, the choice of the PNR detector could be the one described in \cite{eaton_resolution_2023}. Reaching a $3/4$ probability of success has also been found by using non-linear process or ancilla entangled states \cite{PhysRevLett.113.140403,PhysRevA.59.116}.\\

 We can formulate the previous protocol for time-frequency GKP states. If we employ a EPR state  which is less noisy in the temporal and frequency domain, the teleported state is corrected in both temporal and frequency variables at once. It is in contrast with the error correction protocol based on frequency entanglement described in Sec.~\ref{errorentangle}, where we have to repeat twice the same protocol to correct both variables. In the Appendix \ref{correctionGKP}, we tackle the case of the teleportation error correction protocol when the spatial separation of the non-orthogonal qubit state is imperfect, discussing the special case of time-frequency GKP state and the Bell's measurement relying on the spatial separation of the two logical states described in Sec.~\ref{separationGKP}. The imperfect spatial separation affects both the efficiency and the fidelity of the teleportation protocol. The fidelity is always equal to one when the logical states are orthogonal, but this is not the case for non-orthogonal time-frequency qubit states. The use of frequency filters can eliminate detection events caused by imperfect spatial separation, but it comes at the cost of decreased brightness.\\

\section{Conclusion}\label{conclusion}
In this paper, we have analyzed the teleportation-based error correction protocol for two types of frequency qubit states: time-frequency cat states and time-frequency GKP states. This optical scheme has the same goal as a quantum relay, reducing the error rate of wrong detection by decreasing the overlap of the two logical states composing the qubit. We have discussed the experimental realization of Bell's measurement for these two types of qubits. The advantage of discretizing a grid state into a qubit state by combining the even and odd peaks, rather than considering the state as a time-frequency qudit, is convenient because it simplifies the optical implementation of grid state manipulations. When the states are not infinitely frequency narrowed, the Bell's measurement leads to wrong detection, as the spatial separation of the two logical states into two spatial ports is imperfect. This can be corrected by using frequency filters, at the cost of losing single photon detection events. To tackle this issue, the use of frequency resolved detection and the fault-tolerance threshold defined in \cite{fukui_high-threshold_2018} could be valuable for avoiding the use of frequency filters. We have illustrate that our protocol can correct the errors of the qubit composed of two colors, but it could be also done for correcting the temporal error coming from broadening and dispersion, of a qubit composed of two Gaussian centered at two temporal bins (see Eq.~(\ref{timebinqubit})).  In this context, it is important to study and evaluate the overall effectiveness of the teleportation protocol, as well as the final quality of the state being corrected, given the level of accuracy in separating the two logical states.

\section*{ACKNOWLEDGMENT}
 N. Fabre acknowledges useful discussions with Filip Rozp\k{e}dek, Arne Keller and Pérola Milman for the completion of this manuscript.

\appendix

\section{Spatial separation of a two-photon state}\label{Twophoton}

We show in this section that a two-photon state as input can also be separated into the even and the odd components in two distinct spatial ports. We consider an initial separable two photon idea time-frequency GKP state as:
\begin{equation}
\ket{0\tilde{0}}_{aa}=\sum_{n,m\in\mathds{Z}^{2}} c_{2n}\tilde{c}_{2m} \hat{a}^{\dagger}(2n\overline{\omega})\hat{a}^{\dagger}(2m\overline{\omega})\ket{\Omega}.
\end{equation}
The tilde notation is here to indicate that the two states are not identical, one of them can be more noisy compared to the other. After the first beam-splitter and the time-displacement operator, the wave function of the two-photon state is:
\begin{multline}
\frac{1}{2}\sum_{n,m\in\mathds{Z}^{2}} c_{2n}\tilde{c}_{2m}(\hat{a}^{\dagger}_{\tau}(2n\overline{\omega})\hat{a}^{\dagger}_{\tau}(2m\overline{\omega})+\hat{b}^{\dagger}(2n\overline{\omega})\hat{a}^{\dagger}_{\tau}(2m\overline{\omega})\\+\hat{a}^{\dagger}_{\tau}(2n\overline{\omega})\hat{b}^{\dagger}(2m\overline{\omega})+\hat{b}^{\dagger}(2m\overline{\omega})\hat{b}^{\dagger}(2n\overline{\omega}))\ket{\Omega}.
\end{multline}
where $\hat{a}^{\dagger}_{\tau}(2n\overline{\omega})=e^{i2n\overline{\omega}\tau} \hat{a}^{\dagger}(2n\overline{\omega})$. The output wave function after the second beam-splitter is:
\begin{multline}
 \frac{1}{4}\sum_{n,m\in\mathds{Z}^{2}} c_{2n}\tilde{c}_{2m} (\hat{a}^{\dagger}_{\tau}(2n\overline{\omega})+\hat{b}^{\dagger}_{\tau}(2n\overline{\omega}))(\hat{a}^{\dagger}_{\tau}(2m\overline{\omega})+\hat{b}^{\dagger}_{\tau}(2m\overline{\omega}))\\
  +(\hat{a}^{\dagger}_{\tau}(2n\overline{\omega})-\hat{b}^{\dagger}_{\tau}(2n\overline{\omega}))(\hat{a}^{\dagger}_{\tau}(2m\overline{\omega})+\hat{b}^{\dagger}_{\tau}(2m\overline{\omega}))\\
+(\hat{a}^{\dagger}_{\tau}(2n\overline{\omega})+\hat{b}^{\dagger}_{\tau}(2n\overline{\omega}))(\hat{a}^{\dagger}(2m\overline{\omega})-\hat{b}^{\dagger}(2m\overline{\omega}))\\
+ (\hat{a}^{\dagger}(2n\overline{\omega})-\hat{b}^{\dagger}(2n\overline{\omega}))(\hat{a}^{\dagger}(2m\overline{\omega})-\hat{b}^{\dagger}(2m\overline{\omega}))\ket{\Omega}.
\end{multline}
We rearrange and post-select only the coincidence terms:
\begin{multline}
 \frac{1}{4}\sum_{n,m\in\mathds{Z}^{2}} c_{2n}\tilde{c}_{2m}(\hat{a}^{\dagger}_{\tau}(2n\overline{\omega})+\hat{a}^{\dagger}(2n\overline{\omega}))\hat{b}_{\tau}^{\dagger}(2m\overline{\omega})\\
- (\hat{a}^{\dagger}_{\tau}(2n\overline{\omega})+\hat{a}^{\dagger}(2n\overline{\omega}))\hat{b}^{\dagger}(2m\overline{\omega})\\
 +(\hat{a}^{\dagger}_{\tau}(2m\overline{\omega})+\hat{a}^{\dagger}(2m\overline{\omega}))\hat{b}_{\tau}^{\dagger}(2n\overline{\omega})\\
- (\hat{a}^{\dagger}_{\tau}(2m\overline{\omega})+\hat{a}^{\dagger}(2m\overline{\omega}))\hat{b}^{\dagger}(2n\overline{\omega})\ket{\Omega}.
 \end{multline}
Let us first consider the ideal case, where the spectral distribution is a Dirac one $G_{2n}(\omega)=\delta(\omega-2n\overline{\omega})$. We point out that $(\hat{a}^{\dagger}_{\tau}(2n\overline{\omega})+\hat{a}^{\dagger}(2n\overline{\omega}))\ket{0}=(e^{2in\overline{\omega}\tau}+1)\ket{2n\overline{\omega}}$, with $\tau=\pi/\overline{\omega}$, we have $(\hat{a}^{\dagger}_{\tau}(2n\overline{\omega})+\hat{a}^{\dagger}(2n\overline{\omega}))\ket{0}=2\ket{2n\overline{\omega}}$.
 We have also $(\hat{b}_{\tau}^{\dagger}(2m\overline{\omega})-\hat{b}^{\dagger}(2m\overline{\omega})\ket{2n\overline{\omega}}\ket{0}=0$. We can verify that the others terms are zero. It means that there is no coincidence event which is the desired outcome. \\
 
We now rearrange and post-select only the bunching terms:
 \begin{multline}
 \frac{1}{4}\sum_{n,m\in\mathds{Z}^{2}} c_{2n}\tilde{c}_{2m}(\hat{a}^{\dagger}_{\tau}(2n\overline{\omega})\hat{a}^{\dagger}_{\tau}(2m\overline{\omega})+\hat{a}^{\dagger}(2n\overline{\omega})\hat{a}^{\dagger}_{\tau}(2m\overline{\omega})\\
 +\hat{a}^{\dagger}_{\tau}(2n\overline{\omega})\hat{a}^{\dagger}(2m\overline{\omega})+\hat{a}^{\dagger}(2n\overline{\omega})\hat{a}^{\dagger}(2m\overline{\omega})\\
+  \hat{b}^{\dagger}_{\tau}(2n\overline{\omega})\hat{b}^{\dagger}_{\tau}(2m\overline{\omega})-\hat{b}^{\dagger}(2n\overline{\omega})\hat{b}^{\dagger}_{\tau}(2m\overline{\omega})\\
+\hat{b}^{\dagger}_{\tau}(2n\overline{\omega})\hat{b}^{\dagger}(2m\overline{\omega})-\hat{b}^{\dagger}(2n\overline{\omega})\hat{b}^{\dagger}(2m\overline{\omega}))\ket{\Omega}.
 \end{multline}
 In the ideal case, the terms in the spatial port $a$ remains, and the ones in  the spatial port $b$ interfere destructively.\\

For a time-frequency GKP state with a finite bandwidth, there is no longer a perfect destructive (or constructive) interference effect that separates the even and odd components of the comb perfectly. We analyze what happens to the state $\ket{0\tilde{1}}_{aa}$. Post-selecting on the coincidence we have:
\begin{equation}\label{attack}
\ket{0\tilde{1}}_{aa}\rightarrow \ket{0_{G}}_{a}\ket{\tilde{1}_{G}}_{a'}+\ket{0_{E}}_{a'}\ket{\tilde{1}_{E}}_{a}
\end{equation}
where we have defined:
\begin{align}
\ket{0_{G}}_{a}=\frac{N_{e}}{2}\sum_{n\in\mathds{Z}}c_{2n}\int d\omega (e^{i\omega\tau}+1)G_{2n}^{\sigma}(\omega) \ket{\omega}_{a}\\
\ket{\tilde{1}_{G}}_{a'}=\frac{N_{o}}{2}\sum_{n\in\mathds{Z}}c_{2n+1} \int d\omega (e^{i\omega\tau}-1)G_{2m+1}^{\tilde{\sigma}}(\omega) \ket{\omega}_{a'}\\
\ket{\tilde{1}_{E}}_{a}=\frac{N_{o}}{2}\sum_{n\in\mathds{Z}}c_{2n+1} \int d\omega (e^{i\omega\tau}+1)G_{2m+1}^{\tilde{\sigma}}(\omega) \ket{\omega}_{a}\\
\ket{0_{E}}_{a'}=\frac{N_{e}}{2}\sum_{n\in\mathds{Z}}c_{2n}\int d\omega (e^{i\omega\tau}-1)G_{2n}^{\sigma}(\omega) \ket{\omega}_{a'}.
\end{align}
The post-selection on coincidence results in only those events where both the even and odd components are in the correct (designated by $G$) and incorrect (designated by $E$) spatial ports. It should be noted that the output state is no longer in the GKP subspace and results in detection errors. These errors can be corrected through the use of frequency filters or by using frequency-resolved detection and setting a frequency threshold width to accept only certain events \cite{fukui_high-threshold_2018}. Additionally, the imperfect spatial separation described in Eq.~(\ref{attack}) can also be interpreted as an attack to extract some information of the quantum state of interest in a quantum communication protocol.

\section{Teleportation-based error correction protocol with physical time-frequency GKP state}\label{correctionGKP}
In the following, we will employ the GKP coherent picture \cite{seshadreesan_coherent_2021}. The GKP qubit defined by Eq.~(\ref{GKPinput}) is composed of an envelope of width $\kappa$ and each peak has a width of $\sigma$ (resp. $\kappa_{1},\sigma_{1}$), while the frequency widths of the EPR state will be noted $\tilde{\kappa}$ and $\tilde{\sigma}$.\\

As has been noted, when the time-frequency GKP state has a limited bandwidth, the frequency qubit beam-splitter is not able to perfectly separate the odd and even components of the comb. This results in some of the state leaking out of each spatial port, and not conforming to a GKP state. Furthermore, the state in the right port is also distorted. To address these two challenges, it is necessary to employ frequency filters that enable projection back into the GKP subspace and eliminate the undesirable components. Given that the state being teleported and the EPR state have different frequency widths, a frequency filter is placed prior to detection, which establishes the frequency bin and aligns with the reference EPR state. The projector modeling for the frequency filtering process has the following form:
\begin{align}
\hat{\Pi}_{a}=N^{2}_{e}(\sigma,\kappa)\sum_{n\in\mathds{Z}}\tilde{c}_{2n}\int G_{2n}^{\tilde{\sigma}}(\omega)\ket{\omega}\bra{\omega}\\
\hat{\Pi}_{a'}=N^{2}_{o}(\sigma,\kappa)\sum_{n\in\mathds{Z}}\tilde{c}_{2n+1}\int G_{2n+1}^{\tilde{\sigma}}(\omega)\ket{\omega}\bra{\omega}
\end{align}
The normalization constant is found by using $\hat{\Pi}^{2}=\mathds{I}$ and $\text{Tr}(\hat{\Pi}^{2})=1$, $N^{2}_{e}(\tilde{\sigma},\tilde{\kappa})=\sqrt{2\pi}/\tilde{\sigma} \sum_{n}\abs{\tilde{c}_{2n}}^{2}$. In the large comb approximation, we have $N_{e}(\tilde{\sigma},\tilde{\kappa})=N_{o}(\tilde{\sigma},\tilde{\kappa})$. \\

The coincidence destructive measurement is described by the positive operator value measurement:
\begin{equation}
\hat{\Pi}_{a,a'}=\hat{\Pi}_{a}\hat{\Pi}_{a'} \otimes \ket{1}\bra{1}.
\end{equation}
By assuming that the state is pure, the probability of coincidence in the spatial port $a,a'$, $P_{aa'}(0,1)=\text{Tr}(\hat{\Pi}_{a,a'}\ket{\psi}\bra{\psi}\hat{\Pi}_{a,a'}^{\dagger})$ is:
\begin{multline}
P_{aa'}(0,1)=\abs{\frac{1}{2\sqrt{2}}}^{2}\abs{a^{01}_{\sigma\tilde{\sigma}}+b^{01}_{\sigma\tilde{\sigma}}}^{2}\times \\
(\int d\omega (\abs{\alpha}^{2} \abs{\bra{\omega}\ket{\tilde{0}}}^{2}+\abs{\beta}^{2} \abs{\bra{\omega}\ket{\tilde{1}}}^{2}+2\text{Re}(\alpha^{*}\beta \bra{\omega}\ket{\tilde{0}}\bra{\omega}\ket{\tilde{1}}))
\end{multline}
where we have used $\abs{a^{01}_{\sigma\tilde{\sigma}}+b^{01}_{\sigma\tilde{\sigma}}}=\abs{a^{10}_{\sigma\tilde{\sigma}}+b^{10}_{\sigma\tilde{\sigma}}}$ which is shown afterward. For an EPR with a sufficiently narrow distribution, we assume that $\bra{\tilde{0}}\ket{\tilde{1}} =0$, $\int d\omega \abs{\bra{\omega}\ket{\tilde{0}}}^{2}=\int d\omega \abs{\bra{\omega}\ket{\tilde{1}}}^{2}=1$. These two conditions are important, since the probability then does not depend on $\alpha$ and $\beta$,
\begin{equation}
P_{aa'}(0,1)=\abs{\frac{1}{2\sqrt{2}}}^{2}\abs{a^{01}_{\sigma\tilde{\sigma}}+b^{01}_{\sigma\tilde{\sigma}}}^{2}
\end{equation}
since otherwise, information about the quantum state could be extracted during the measurement.\\

The wavefunction of the state after the detection is: $\ket{\psi}_{c}=\hat{\Pi}_{i,j}\ket{\psi}/\text{Tr}(\hat{\Pi}_{i,j}\ket{\psi}\bra{\psi}\hat{\Pi}_{i,j}^{\dagger})$, where $i,j=a,b; a',b'$. When coincidence is detected at the spatial port $a$ and $a'$, the post-selected state is:
\begin{multline}\label{postselectedstatewithout}
\ket{\psi}_{c}=\frac{1}{\abs{a^{01}_{\sigma\tilde{\sigma}}+b^{01}_{\sigma\tilde{\sigma}}}} (\alpha(a^{1,\tilde{\sigma}}_{0, \sigma}+b^{1,\sigma}_{0,\tilde{\sigma}}) \ket{\tilde{0}}_{c}\\+\beta(a^{1,\sigma}_{0, \tilde{\sigma}}
+b^{1,\tilde{\sigma}}_{0,\sigma}) \ket{\tilde{1}}_{c}).
\end{multline}
 The expression is similar for the other coincidences events in the other spatial ports, and extra Pauli operations have to be performed. The coefficients $a^{10}_{\tilde{\sigma}\sigma}$ and $b^{10}_{\sigma \tilde{\sigma}}$ have the expression:
\begin{multline}
a^{10}_{\tilde{\sigma}\sigma}=\frac{1}{4} N_{e}(\sigma,\kappa)N_{o}(\tilde{\sigma},\tilde{\kappa})N_{e}(\tilde{\sigma},\tilde{\kappa})N_{o}(\tilde{\sigma},\tilde{\kappa})\\
\times \sum_{n,m,k,k'} c_{2n}\tilde{c}_{2m+1}\tilde{c}_{2k}\tilde{c}_{2k'+1}\int d\omega (e^{i\omega\tau}+1)G_{2n}^{\sigma}(\omega)G_{2k}^{\tilde{\sigma}}(\omega) \\
\times \int d\omega (e^{i\omega\tau}-1)G_{2m+1}^{\tilde{\sigma}}(\omega) )G_{2k'+1}^{\tilde{\sigma}}(\omega)
\end{multline}
\begin{multline}
b^{10}_{\sigma \tilde{\sigma}}=\frac{1}{4} N_{e}(\tilde{\sigma},\tilde{\kappa})N_{o}(\sigma,\kappa)N_{e}(\tilde{\sigma},\tilde{\kappa})N_{o}(\tilde{\sigma},\tilde{\kappa})\\
\times \sum_{n,m,k,k'} \tilde{c}_{2n}c_{2m+1}\tilde{c}_{2k}\tilde{c}_{2k'+1}\int d\omega (e^{i\omega\tau}+1)G_{2m+1}^{\sigma}(\omega)G_{2k}^{\tilde{\sigma}}(\omega)\\
\times  \int d\omega (e^{i\omega\tau}-1)G_{2n}^{\tilde{\sigma}}(\omega) )G_{2k'+1}^{\tilde{\sigma}}(\omega).
\end{multline}
The $b$ coefficient contains odd (resp. even) terms in a spatial port where the frequency filters is centered at even (resp. odd) frequencies. In the ideal case, namely if both the EPR and the teleported state are ideal time-frequency GKP state we remind that $a^{01}_{\sigma\tilde{\sigma}}=2$ and $b^{01}_{\sigma\tilde{\sigma}}=0$. After evaluation of the integrals, and assuming that $2k=2n$, meaning that the temporal spreading only reach the next bin, we find that:
\begin{multline}\label{definitionaandb}
a^{10}_{\tilde{\sigma}\sigma}=\frac{\sqrt{\tilde{\sigma}\tilde{\sigma}}\sqrt{\sigma\tilde{\sigma}}}{4\sqrt{\tilde{\sigma}^{2}+\sigma^{2}}\sqrt{\tilde{\sigma}^{2}+\tilde{\sigma}^{2}}}(e^{-\frac{\pi^{2}\alpha^{2}}{2\overline{\omega}^{2}}}+1)(-e^{-\frac{\pi^{2}\tilde{\alpha}^{2}}{2\overline{\omega}^{2}}}-1)
\end{multline}
\begin{multline}
b^{10}_{\sigma \tilde{\sigma}}=\frac{\sqrt{\tilde{\sigma}\tilde{\sigma}}\sqrt{\sigma\tilde{\sigma}}}{4\sqrt{\tilde{\sigma}^{2}+\sigma^{2}}\sqrt{\tilde{\sigma}^{2}+\tilde{\sigma}^{2}}}\frac{\sum_{n,m} c_{2m+1}c_{2m}c_{2n}c_{2n+1}}{\sum_{n,m} \abs{c_{2m+1}}^{2}\abs{c_{2n}}^{2}}\\
\times (e^{-\frac{\pi^{2}\alpha^{2}}{2\overline{\omega}^{2}}} e^{-\frac{\overline{\omega}^{2}}{2(\tilde{\sigma}^{2}+\tilde{\sigma}^{2})}} e^{i\frac{\pi \tilde{\sigma}^{2}}{(\sigma^{2}+\tilde{\sigma}^{2})}}+1)\\
\times (e^{-\frac{\pi^{2}\alpha^{2}}{2\overline{\omega}^{2}}}e^{-\frac{\overline{\omega}^{2}}{2(\tilde{\sigma}^{2}+\tilde{\sigma}^{2})}} e^{i\frac{\pi \tilde{\sigma}^{2}}{(\sigma^{2}+\tilde{\sigma}^{2})}}-1)
\end{multline}
where we have defined that $\alpha^{2}=\sigma^{2}\tilde{\sigma}^{2}/(\sigma^{2}+\tilde{\sigma}^{2})$. While $a^{10}_{\tilde{\sigma}\sigma}$ is real, $b^{10}_{\sigma \tilde{\sigma}}$ is a complex quantity. From these expressions, we point out that $a^{1,\tilde{\sigma}}_{0, \sigma}=a^{0,\tilde{\sigma}}_{1, \sigma}$ and $b^{1,\sigma}_{0,\tilde{\sigma}}=b^{0,\sigma}_{1,\tilde{\sigma}}$. When the width of the frequency filter is $\tilde{\sigma}\rightarrow 0$, we have $a^{01}_{\sigma\tilde{\sigma}}=2$ and $b^{01}_{\sigma\tilde{\sigma}}=0$ which is as the ideal case and thus lead to a high fidelity of the state, but at the cost of losing many photons.   As the imperfect spatial separation of the two logical states leads to a decreasing of the fidelity, it is a reminiscent fact coming that the state possesses continuous variables.\\

\bibliography{biblio}

\begin{thebibliography}{64}%
\makeatletter
\providecommand \@ifxundefined [1]{%
 \@ifx{#1\undefined}
}%
\providecommand \@ifnum [1]{%
 \ifnum #1\expandafter \@firstoftwo
 \else \expandafter \@secondoftwo
 \fi
}%
\providecommand \@ifx [1]{%
 \ifx #1\expandafter \@firstoftwo
 \else \expandafter \@secondoftwo
 \fi
}%
\providecommand \natexlab [1]{#1}%
\providecommand \enquote  [1]{``#1''}%
\providecommand \bibnamefont  [1]{#1}%
\providecommand \bibfnamefont [1]{#1}%
\providecommand \citenamefont [1]{#1}%
\providecommand \href@noop [0]{\@secondoftwo}%
\providecommand \href [0]{\begingroup \@sanitize@url \@href}%
\providecommand \@href[1]{\@@startlink{#1}\@@href}%
\providecommand \@@href[1]{\endgroup#1\@@endlink}%
\providecommand \@sanitize@url [0]{\catcode `\\12\catcode `\$12\catcode
  `\&12\catcode `\#12\catcode `\^12\catcode `\_12\catcode `\%12\relax}%
\providecommand \@@startlink[1]{}%
\providecommand \@@endlink[0]{}%
\providecommand \url  [0]{\begingroup\@sanitize@url \@url }%
\providecommand \@url [1]{\endgroup\@href {#1}{\urlprefix }}%
\providecommand \urlprefix  [0]{URL }%
\providecommand \Eprint [0]{\href }%
\providecommand \doibase [0]{https://doi.org/}%
\providecommand \selectlanguage [0]{\@gobble}%
\providecommand \bibinfo  [0]{\@secondoftwo}%
\providecommand \bibfield  [0]{\@secondoftwo}%
\providecommand \translation [1]{[#1]}%
\providecommand \BibitemOpen [0]{}%
\providecommand \bibitemStop [0]{}%
\providecommand \bibitemNoStop [0]{.\EOS\space}%
\providecommand \EOS [0]{\spacefactor3000\relax}%
\providecommand \BibitemShut  [1]{\csname bibitem#1\endcsname}%
\let\auto@bib@innerbib\@empty
\bibitem [{\citenamefont {Fabre}\ \emph
  {et~al.}(2020{\natexlab{a}})\citenamefont {Fabre}, \citenamefont {Maltese},
  \citenamefont {Appas}, \citenamefont {Felicetti}, \citenamefont {Ketterer},
  \citenamefont {Keller}, \citenamefont {Coudreau}, \citenamefont {Baboux},
  \citenamefont {Amanti}, \citenamefont {Ducci},\ and\ \citenamefont
  {Milman}}]{fabre_generation_2020}%
  \BibitemOpen
  \bibfield  {author} {\bibinfo {author} {\bibfnamefont {N.}~\bibnamefont
  {Fabre}}, \bibinfo {author} {\bibfnamefont {G.}~\bibnamefont {Maltese}},
  \bibinfo {author} {\bibfnamefont {F.}~\bibnamefont {Appas}}, \bibinfo
  {author} {\bibfnamefont {S.}~\bibnamefont {Felicetti}}, \bibinfo {author}
  {\bibfnamefont {A.}~\bibnamefont {Ketterer}}, \bibinfo {author}
  {\bibfnamefont {A.}~\bibnamefont {Keller}}, \bibinfo {author} {\bibfnamefont
  {T.}~\bibnamefont {Coudreau}}, \bibinfo {author} {\bibfnamefont
  {F.}~\bibnamefont {Baboux}}, \bibinfo {author} {\bibfnamefont {M.~I.}\
  \bibnamefont {Amanti}}, \bibinfo {author} {\bibfnamefont {S.}~\bibnamefont
  {Ducci}},\ and\ \bibinfo {author} {\bibfnamefont {P.}~\bibnamefont
  {Milman}},\ }\bibfield  {title} {\bibinfo {title} {Generation of a
  time-frequency grid state with integrated biphoton frequency combs},\ }\href
  {https://doi.org/10.1103/PhysRevA.102.012607} {\bibfield  {journal} {\bibinfo
   {journal} {Phys. Rev. A}\ }\textbf {\bibinfo {volume} {102}},\ \bibinfo
  {pages} {012607} (\bibinfo {year} {2020}{\natexlab{a}})},\ \bibinfo {note}
  {publisher: American Physical Society}\BibitemShut {NoStop}%
\bibitem [{\citenamefont {Fabre}(2020)}]{fabre:tel-03191301}%
  \BibitemOpen
  \bibfield  {author} {\bibinfo {author} {\bibfnamefont {N.}~\bibnamefont
  {Fabre}},\ }\emph {\bibinfo {title} {{Quantum information in time-frequency
  continuous variables}}},\ \href
  {https://tel.archives-ouvertes.fr/tel-03191301} {\bibinfo {type} {Quantum
  information in time-frequency continuous variables}},\ \bibinfo  {school}
  {{Universit{\'e} de Paris}} (\bibinfo {year} {2020})\BibitemShut {NoStop}%
\bibitem [{\citenamefont {Tasca}\ \emph {et~al.}(2011)\citenamefont {Tasca},
  \citenamefont {Gomes}, \citenamefont {Toscano}, \citenamefont {Ribeiro},\
  and\ \citenamefont {Walborn}}]{tasca_continuous_2011}%
  \BibitemOpen
  \bibfield  {author} {\bibinfo {author} {\bibfnamefont {D.~S.}\ \bibnamefont
  {Tasca}}, \bibinfo {author} {\bibfnamefont {R.~M.}\ \bibnamefont {Gomes}},
  \bibinfo {author} {\bibfnamefont {F.}~\bibnamefont {Toscano}}, \bibinfo
  {author} {\bibfnamefont {P.~H.~S.}\ \bibnamefont {Ribeiro}},\ and\ \bibinfo
  {author} {\bibfnamefont {S.~P.}\ \bibnamefont {Walborn}},\ }\bibfield
  {title} {\bibinfo {title} {Continuous variable quantum computation with
  spatial degrees of freedom of photons},\ }\href
  {https://doi.org/10.1103/PhysRevA.83.052325} {\bibfield  {journal} {\bibinfo
  {journal} {Phys. Rev. A}\ }\textbf {\bibinfo {volume} {83}},\ \bibinfo
  {pages} {052325} (\bibinfo {year} {2011})},\ \Eprint
  {https://arxiv.org/abs/1106.3049} {1106.3049} \BibitemShut {NoStop}%
\bibitem [{\citenamefont {Fabre}\ \emph {et~al.}(2022)\citenamefont {Fabre},
  \citenamefont {Keller},\ and\ \citenamefont {Milman}}]{fabre_time_2022}%
  \BibitemOpen
  \bibfield  {author} {\bibinfo {author} {\bibfnamefont {N.}~\bibnamefont
  {Fabre}}, \bibinfo {author} {\bibfnamefont {A.}~\bibnamefont {Keller}},\ and\
  \bibinfo {author} {\bibfnamefont {P.}~\bibnamefont {Milman}},\ }\bibfield
  {title} {\bibinfo {title} {Time and frequency as quantum continuous
  variables},\ }\href {https://doi.org/10.1103/PhysRevA.105.052429} {\bibfield
  {journal} {\bibinfo  {journal} {Phys. Rev. A}\ }\textbf {\bibinfo {volume}
  {105}},\ \bibinfo {pages} {052429} (\bibinfo {year} {2022})},\ \bibinfo
  {note} {publisher: American Physical Society}\BibitemShut {NoStop}%
\bibitem [{\citenamefont {Tsang}\ \emph {et~al.}(2016)\citenamefont {Tsang},
  \citenamefont {Nair},\ and\ \citenamefont {Lu}}]{PhysRevX.6.031033}%
  \BibitemOpen
  \bibfield  {author} {\bibinfo {author} {\bibfnamefont {M.}~\bibnamefont
  {Tsang}}, \bibinfo {author} {\bibfnamefont {R.}~\bibnamefont {Nair}},\ and\
  \bibinfo {author} {\bibfnamefont {X.-M.}\ \bibnamefont {Lu}},\ }\bibfield
  {title} {\bibinfo {title} {Quantum theory of superresolution for two
  incoherent optical point sources},\ }\href
  {https://doi.org/10.1103/PhysRevX.6.031033} {\bibfield  {journal} {\bibinfo
  {journal} {Phys. Rev. X}\ }\textbf {\bibinfo {volume} {6}},\ \bibinfo {pages}
  {031033} (\bibinfo {year} {2016})}\BibitemShut {NoStop}%
\bibitem [{\citenamefont {Cochrane}\ \emph {et~al.}(1999)\citenamefont
  {Cochrane}, \citenamefont {Milburn},\ and\ \citenamefont
  {Munro}}]{cochrane_macroscopically_1999}%
  \BibitemOpen
  \bibfield  {author} {\bibinfo {author} {\bibfnamefont {P.~T.}\ \bibnamefont
  {Cochrane}}, \bibinfo {author} {\bibfnamefont {G.~J.}\ \bibnamefont
  {Milburn}},\ and\ \bibinfo {author} {\bibfnamefont {W.~J.}\ \bibnamefont
  {Munro}},\ }\bibfield  {title} {\bibinfo {title} {Macroscopically distinct
  quantum superposition states as a bosonic code for amplitude damping},\
  }\href {https://doi.org/10.1103/PhysRevA.59.2631} {\bibfield  {journal}
  {\bibinfo  {journal} {Phys. Rev. A}\ }\textbf {\bibinfo {volume} {59}},\
  \bibinfo {pages} {2631} (\bibinfo {year} {1999})},\ \Eprint
  {https://arxiv.org/abs/quant-ph/9809037} {quant-ph/9809037} \BibitemShut
  {NoStop}%
\bibitem [{\citenamefont {Guillaud}\ and\ \citenamefont
  {Mirrahimi}(2019)}]{PhysRevX.9.041053}%
  \BibitemOpen
  \bibfield  {author} {\bibinfo {author} {\bibfnamefont {J.}~\bibnamefont
  {Guillaud}}\ and\ \bibinfo {author} {\bibfnamefont {M.}~\bibnamefont
  {Mirrahimi}},\ }\bibfield  {title} {\bibinfo {title} {Repetition cat qubits
  for fault-tolerant quantum computation},\ }\href
  {https://doi.org/10.1103/PhysRevX.9.041053} {\bibfield  {journal} {\bibinfo
  {journal} {Phys. Rev. X}\ }\textbf {\bibinfo {volume} {9}},\ \bibinfo {pages}
  {041053} (\bibinfo {year} {2019})}\BibitemShut {NoStop}%
\bibitem [{\citenamefont {Albert}\ \emph {et~al.}(2019)\citenamefont {Albert},
  \citenamefont {Mundhada}, \citenamefont {Grimm}, \citenamefont {Touzard},
  \citenamefont {Devoret},\ and\ \citenamefont {Jiang}}]{Albert_2019}%
  \BibitemOpen
  \bibfield  {author} {\bibinfo {author} {\bibfnamefont {V.~V.}\ \bibnamefont
  {Albert}}, \bibinfo {author} {\bibfnamefont {S.~O.}\ \bibnamefont
  {Mundhada}}, \bibinfo {author} {\bibfnamefont {A.}~\bibnamefont {Grimm}},
  \bibinfo {author} {\bibfnamefont {S.}~\bibnamefont {Touzard}}, \bibinfo
  {author} {\bibfnamefont {M.~H.}\ \bibnamefont {Devoret}},\ and\ \bibinfo
  {author} {\bibfnamefont {L.}~\bibnamefont {Jiang}},\ }\bibfield  {title}
  {\bibinfo {title} {Pair-cat codes: autonomous error-correction with low-order
  nonlinearity},\ }\href {https://doi.org/10.1088/2058-9565/ab1e69} {\bibfield
  {journal} {\bibinfo  {journal} {Quantum Science and Technology}\ }\textbf
  {\bibinfo {volume} {4}},\ \bibinfo {pages} {035007} (\bibinfo {year}
  {2019})}\BibitemShut {NoStop}%
\bibitem [{\citenamefont {Gottesman}\ \emph {et~al.}(2001)\citenamefont
  {Gottesman}, \citenamefont {Kitaev},\ and\ \citenamefont
  {Preskill}}]{gottesman_encoding_2001}%
  \BibitemOpen
  \bibfield  {author} {\bibinfo {author} {\bibfnamefont {D.}~\bibnamefont
  {Gottesman}}, \bibinfo {author} {\bibfnamefont {A.}~\bibnamefont {Kitaev}},\
  and\ \bibinfo {author} {\bibfnamefont {J.}~\bibnamefont {Preskill}},\
  }\bibfield  {title} {\bibinfo {title} {Encoding a qubit in an oscillator},\
  }\href {https://doi.org/10.1103/PhysRevA.64.012310} {\bibfield  {journal}
  {\bibinfo  {journal} {Phys. Rev. A}\ }\textbf {\bibinfo {volume} {64}},\
  \bibinfo {pages} {012310} (\bibinfo {year} {2001})}\BibitemShut {NoStop}%
\bibitem [{\citenamefont {Flühmann}\ \emph {et~al.}(2019)\citenamefont
  {Flühmann}, \citenamefont {Nguyen}, \citenamefont {Marinelli}, \citenamefont
  {Negnevitsky}, \citenamefont {Mehta},\ and\ \citenamefont
  {Home}}]{fluhmann_encoding_2019}%
  \BibitemOpen
  \bibfield  {author} {\bibinfo {author} {\bibfnamefont {C.}~\bibnamefont
  {Flühmann}}, \bibinfo {author} {\bibfnamefont {T.~L.}\ \bibnamefont
  {Nguyen}}, \bibinfo {author} {\bibfnamefont {M.}~\bibnamefont {Marinelli}},
  \bibinfo {author} {\bibfnamefont {V.}~\bibnamefont {Negnevitsky}}, \bibinfo
  {author} {\bibfnamefont {K.}~\bibnamefont {Mehta}},\ and\ \bibinfo {author}
  {\bibfnamefont {J.~P.}\ \bibnamefont {Home}},\ }\bibfield  {title} {\bibinfo
  {title} {Encoding a qubit in a trapped-ion mechanical oscillator},\ }\href
  {https://doi.org/10.1038/s41586-019-0960-6} {\bibfield  {journal} {\bibinfo
  {journal} {Nature}\ }\textbf {\bibinfo {volume} {566}},\ \bibinfo {pages}
  {513} (\bibinfo {year} {2019})}\BibitemShut {NoStop}%
\bibitem [{\citenamefont {Vuillot}\ \emph {et~al.}(2019)\citenamefont
  {Vuillot}, \citenamefont {Asasi}, \citenamefont {Wang}, \citenamefont
  {Pryadko},\ and\ \citenamefont {Terhal}}]{PhysRevA.99.032344}%
  \BibitemOpen
  \bibfield  {author} {\bibinfo {author} {\bibfnamefont {C.}~\bibnamefont
  {Vuillot}}, \bibinfo {author} {\bibfnamefont {H.}~\bibnamefont {Asasi}},
  \bibinfo {author} {\bibfnamefont {Y.}~\bibnamefont {Wang}}, \bibinfo {author}
  {\bibfnamefont {L.~P.}\ \bibnamefont {Pryadko}},\ and\ \bibinfo {author}
  {\bibfnamefont {B.~M.}\ \bibnamefont {Terhal}},\ }\bibfield  {title}
  {\bibinfo {title} {Quantum error correction with the toric
  gottesman-kitaev-preskill code},\ }\href
  {https://doi.org/10.1103/PhysRevA.99.032344} {\bibfield  {journal} {\bibinfo
  {journal} {Phys. Rev. A}\ }\textbf {\bibinfo {volume} {99}},\ \bibinfo
  {pages} {032344} (\bibinfo {year} {2019})}\BibitemShut {NoStop}%
\bibitem [{\citenamefont {Campagne-Ibarcq}\ \emph {et~al.}(2020)\citenamefont
  {Campagne-Ibarcq}, \citenamefont {Eickbusch}, \citenamefont {Touzard},
  \citenamefont {Zalys-Geller}, \citenamefont {Frattini}, \citenamefont
  {Sivak}, \citenamefont {Reinhold}, \citenamefont {Puri}, \citenamefont
  {Shankar}, \citenamefont {Schoelkopf}, \citenamefont {Frunzio}, \citenamefont
  {Mirrahimi},\ and\ \citenamefont {Devoret}}]{campagne-ibarcq_quantum_2020}%
  \BibitemOpen
  \bibfield  {author} {\bibinfo {author} {\bibfnamefont {P.}~\bibnamefont
  {Campagne-Ibarcq}}, \bibinfo {author} {\bibfnamefont {A.}~\bibnamefont
  {Eickbusch}}, \bibinfo {author} {\bibfnamefont {S.}~\bibnamefont {Touzard}},
  \bibinfo {author} {\bibfnamefont {E.}~\bibnamefont {Zalys-Geller}}, \bibinfo
  {author} {\bibfnamefont {N.~E.}\ \bibnamefont {Frattini}}, \bibinfo {author}
  {\bibfnamefont {V.~V.}\ \bibnamefont {Sivak}}, \bibinfo {author}
  {\bibfnamefont {P.}~\bibnamefont {Reinhold}}, \bibinfo {author}
  {\bibfnamefont {S.}~\bibnamefont {Puri}}, \bibinfo {author} {\bibfnamefont
  {S.}~\bibnamefont {Shankar}}, \bibinfo {author} {\bibfnamefont {R.~J.}\
  \bibnamefont {Schoelkopf}}, \bibinfo {author} {\bibfnamefont
  {L.}~\bibnamefont {Frunzio}}, \bibinfo {author} {\bibfnamefont
  {M.}~\bibnamefont {Mirrahimi}},\ and\ \bibinfo {author} {\bibfnamefont
  {M.~H.}\ \bibnamefont {Devoret}},\ }\bibfield  {title} {\bibinfo {title}
  {Quantum error correction of a qubit encoded in grid states of an
  oscillator},\ }\href {https://doi.org/10.1038/s41586-020-2603-3} {\bibfield
  {journal} {\bibinfo  {journal} {Nature}\ }\textbf {\bibinfo {volume} {584}},\
  \bibinfo {pages} {368} (\bibinfo {year} {2020})}\BibitemShut {NoStop}%
\bibitem [{\citenamefont {Calcluth}\ \emph {et~al.}(2022)\citenamefont
  {Calcluth}, \citenamefont {Ferraro},\ and\ \citenamefont
  {Ferrini}}]{https://doi.org/10.48550/arxiv.2205.09781}%
  \BibitemOpen
  \bibfield  {author} {\bibinfo {author} {\bibfnamefont {C.}~\bibnamefont
  {Calcluth}}, \bibinfo {author} {\bibfnamefont {A.}~\bibnamefont {Ferraro}},\
  and\ \bibinfo {author} {\bibfnamefont {G.}~\bibnamefont {Ferrini}},\ }\href
  {https://doi.org/10.48550/ARXIV.2205.09781} {\bibinfo {title} {The vacuum
  provides quantum advantage to otherwise simulatable architectures}} (\bibinfo
  {year} {2022})\BibitemShut {NoStop}%
\bibitem [{\citenamefont {Hastrup}\ \emph {et~al.}(2021)\citenamefont
  {Hastrup}, \citenamefont {Larsen}, \citenamefont {Neergaard-Nielsen},
  \citenamefont {Menicucci},\ and\ \citenamefont
  {Andersen}}]{PhysRevA.103.032409}%
  \BibitemOpen
  \bibfield  {author} {\bibinfo {author} {\bibfnamefont {J.}~\bibnamefont
  {Hastrup}}, \bibinfo {author} {\bibfnamefont {M.~V.}\ \bibnamefont {Larsen}},
  \bibinfo {author} {\bibfnamefont {J.~S.}\ \bibnamefont {Neergaard-Nielsen}},
  \bibinfo {author} {\bibfnamefont {N.~C.}\ \bibnamefont {Menicucci}},\ and\
  \bibinfo {author} {\bibfnamefont {U.~L.}\ \bibnamefont {Andersen}},\
  }\bibfield  {title} {\bibinfo {title} {Unsuitability of cubic phase gates for
  non-clifford operations on gottesman-kitaev-preskill states},\ }\href
  {https://doi.org/10.1103/PhysRevA.103.032409} {\bibfield  {journal} {\bibinfo
   {journal} {Phys. Rev. A}\ }\textbf {\bibinfo {volume} {103}},\ \bibinfo
  {pages} {032409} (\bibinfo {year} {2021})}\BibitemShut {NoStop}%
\bibitem [{\citenamefont {Noh}\ \emph {et~al.}(2019)\citenamefont {Noh},
  \citenamefont {Albert},\ and\ \citenamefont {Jiang}}]{8482307}%
  \BibitemOpen
  \bibfield  {author} {\bibinfo {author} {\bibfnamefont {K.}~\bibnamefont
  {Noh}}, \bibinfo {author} {\bibfnamefont {V.~V.}\ \bibnamefont {Albert}},\
  and\ \bibinfo {author} {\bibfnamefont {L.}~\bibnamefont {Jiang}},\ }\bibfield
   {title} {\bibinfo {title} {Quantum capacity bounds of gaussian thermal loss
  channels and achievable rates with gottesman-kitaev-preskill codes},\ }\href
  {https://doi.org/10.1109/TIT.2018.2873764} {\bibfield  {journal} {\bibinfo
  {journal} {IEEE Transactions on Information Theory}\ }\textbf {\bibinfo
  {volume} {65}},\ \bibinfo {pages} {2563} (\bibinfo {year}
  {2019})}\BibitemShut {NoStop}%
\bibitem [{\citenamefont {Michael}\ \emph {et~al.}(2016)\citenamefont
  {Michael}, \citenamefont {Silveri}, \citenamefont {Brierley}, \citenamefont
  {Albert}, \citenamefont {Salmilehto}, \citenamefont {Jiang},\ and\
  \citenamefont {Girvin}}]{michael_new_2016}%
  \BibitemOpen
  \bibfield  {author} {\bibinfo {author} {\bibfnamefont {M.~H.}\ \bibnamefont
  {Michael}}, \bibinfo {author} {\bibfnamefont {M.}~\bibnamefont {Silveri}},
  \bibinfo {author} {\bibfnamefont {R.~T.}\ \bibnamefont {Brierley}}, \bibinfo
  {author} {\bibfnamefont {V.~V.}\ \bibnamefont {Albert}}, \bibinfo {author}
  {\bibfnamefont {J.}~\bibnamefont {Salmilehto}}, \bibinfo {author}
  {\bibfnamefont {L.}~\bibnamefont {Jiang}},\ and\ \bibinfo {author}
  {\bibfnamefont {S.~M.}\ \bibnamefont {Girvin}},\ }\bibfield  {title}
  {\bibinfo {title} {New class of quantum error-correcting codes for a bosonic
  mode},\ }\href {https://doi.org/10.1103/PhysRevX.6.031006} {\bibfield
  {journal} {\bibinfo  {journal} {Phys. Rev. X}\ }\textbf {\bibinfo {volume}
  {6}},\ \bibinfo {pages} {031006} (\bibinfo {year} {2016})},\ \Eprint
  {https://arxiv.org/abs/1602.00008} {1602.00008} \BibitemShut {NoStop}%
\bibitem [{\citenamefont {Hu}\ \emph {et~al.}(2019)\citenamefont {Hu},
  \citenamefont {Ma}, \citenamefont {Cai}, \citenamefont {Mu}, \citenamefont
  {Xu}, \citenamefont {Wang}, \citenamefont {Wu}, \citenamefont {Wang},
  \citenamefont {Song}, \citenamefont {Zou}, \citenamefont {Girvin},
  \citenamefont {Duan},\ and\ \citenamefont {Sun}}]{hu_quantum_2019}%
  \BibitemOpen
  \bibfield  {author} {\bibinfo {author} {\bibfnamefont {L.}~\bibnamefont
  {Hu}}, \bibinfo {author} {\bibfnamefont {Y.}~\bibnamefont {Ma}}, \bibinfo
  {author} {\bibfnamefont {W.}~\bibnamefont {Cai}}, \bibinfo {author}
  {\bibfnamefont {X.}~\bibnamefont {Mu}}, \bibinfo {author} {\bibfnamefont
  {Y.}~\bibnamefont {Xu}}, \bibinfo {author} {\bibfnamefont {W.}~\bibnamefont
  {Wang}}, \bibinfo {author} {\bibfnamefont {Y.}~\bibnamefont {Wu}}, \bibinfo
  {author} {\bibfnamefont {H.}~\bibnamefont {Wang}}, \bibinfo {author}
  {\bibfnamefont {Y.~P.}\ \bibnamefont {Song}}, \bibinfo {author}
  {\bibfnamefont {C.-L.}\ \bibnamefont {Zou}}, \bibinfo {author} {\bibfnamefont
  {S.~M.}\ \bibnamefont {Girvin}}, \bibinfo {author} {\bibfnamefont {L.-M.}\
  \bibnamefont {Duan}},\ and\ \bibinfo {author} {\bibfnamefont
  {L.}~\bibnamefont {Sun}},\ }\bibfield  {title} {\bibinfo {title} {Quantum
  error correction and universal gate set operation on a binomial bosonic
  logical qubit},\ }\href {https://doi.org/10.1038/s41567-018-0414-3}
  {\bibfield  {journal} {\bibinfo  {journal} {Nature Physics}\ }\textbf
  {\bibinfo {volume} {15}},\ \bibinfo {pages} {503} (\bibinfo {year}
  {2019})}\BibitemShut {NoStop}%
\bibitem [{\citenamefont {Chamberland}\ \emph {et~al.}(2022)\citenamefont
  {Chamberland}, \citenamefont {Noh}, \citenamefont {Arrangoiz-Arriola},
  \citenamefont {Campbell}, \citenamefont {Hann}, \citenamefont {Iverson},
  \citenamefont {Putterman}, \citenamefont {Bohdanowicz}, \citenamefont
  {Flammia}, \citenamefont {Keller}, \citenamefont {Refael}, \citenamefont
  {Preskill}, \citenamefont {Jiang}, \citenamefont {Safavi-Naeini},
  \citenamefont {Painter},\ and\ \citenamefont
  {Brand\~ao}}]{PRXQuantum.3.010329}%
  \BibitemOpen
  \bibfield  {author} {\bibinfo {author} {\bibfnamefont {C.}~\bibnamefont
  {Chamberland}}, \bibinfo {author} {\bibfnamefont {K.}~\bibnamefont {Noh}},
  \bibinfo {author} {\bibfnamefont {P.}~\bibnamefont {Arrangoiz-Arriola}},
  \bibinfo {author} {\bibfnamefont {E.~T.}\ \bibnamefont {Campbell}}, \bibinfo
  {author} {\bibfnamefont {C.~T.}\ \bibnamefont {Hann}}, \bibinfo {author}
  {\bibfnamefont {J.}~\bibnamefont {Iverson}}, \bibinfo {author} {\bibfnamefont
  {H.}~\bibnamefont {Putterman}}, \bibinfo {author} {\bibfnamefont {T.~C.}\
  \bibnamefont {Bohdanowicz}}, \bibinfo {author} {\bibfnamefont {S.~T.}\
  \bibnamefont {Flammia}}, \bibinfo {author} {\bibfnamefont {A.}~\bibnamefont
  {Keller}}, \bibinfo {author} {\bibfnamefont {G.}~\bibnamefont {Refael}},
  \bibinfo {author} {\bibfnamefont {J.}~\bibnamefont {Preskill}}, \bibinfo
  {author} {\bibfnamefont {L.}~\bibnamefont {Jiang}}, \bibinfo {author}
  {\bibfnamefont {A.~H.}\ \bibnamefont {Safavi-Naeini}}, \bibinfo {author}
  {\bibfnamefont {O.}~\bibnamefont {Painter}},\ and\ \bibinfo {author}
  {\bibfnamefont {F.~G.}\ \bibnamefont {Brand\~ao}},\ }\bibfield  {title}
  {\bibinfo {title} {Building a fault-tolerant quantum computer using
  concatenated cat codes},\ }\href
  {https://doi.org/10.1103/PRXQuantum.3.010329} {\bibfield  {journal} {\bibinfo
   {journal} {PRX Quantum}\ }\textbf {\bibinfo {volume} {3}},\ \bibinfo {pages}
  {010329} (\bibinfo {year} {2022})}\BibitemShut {NoStop}%
\bibitem [{\citenamefont {Baragiola}\ \emph {et~al.}(2019)\citenamefont
  {Baragiola}, \citenamefont {Pantaleoni}, \citenamefont {Alexander},
  \citenamefont {Karanjai},\ and\ \citenamefont
  {Menicucci}}]{PhysRevLett.123.200502}%
  \BibitemOpen
  \bibfield  {author} {\bibinfo {author} {\bibfnamefont {B.~Q.}\ \bibnamefont
  {Baragiola}}, \bibinfo {author} {\bibfnamefont {G.}~\bibnamefont
  {Pantaleoni}}, \bibinfo {author} {\bibfnamefont {R.~N.}\ \bibnamefont
  {Alexander}}, \bibinfo {author} {\bibfnamefont {A.}~\bibnamefont
  {Karanjai}},\ and\ \bibinfo {author} {\bibfnamefont {N.~C.}\ \bibnamefont
  {Menicucci}},\ }\bibfield  {title} {\bibinfo {title} {All-gaussian
  universality and fault tolerance with the gottesman-kitaev-preskill code},\
  }\href {https://doi.org/10.1103/PhysRevLett.123.200502} {\bibfield  {journal}
  {\bibinfo  {journal} {Phys. Rev. Lett.}\ }\textbf {\bibinfo {volume} {123}},\
  \bibinfo {pages} {200502} (\bibinfo {year} {2019})}\BibitemShut {NoStop}%
\bibitem [{\citenamefont {Bourassa}\ \emph {et~al.}(2021)\citenamefont
  {Bourassa}, \citenamefont {Alexander}, \citenamefont {Vasmer}, \citenamefont
  {Patil}, \citenamefont {Tzitrin}, \citenamefont {Matsuura}, \citenamefont
  {Su}, \citenamefont {Baragiola}, \citenamefont {Guha}, \citenamefont
  {Dauphinais}, \citenamefont {Sabapathy}, \citenamefont {Menicucci},\ and\
  \citenamefont {Dhand}}]{bourassa_blueprint_2021}%
  \BibitemOpen
  \bibfield  {author} {\bibinfo {author} {\bibfnamefont {J.~E.}\ \bibnamefont
  {Bourassa}}, \bibinfo {author} {\bibfnamefont {R.~N.}\ \bibnamefont
  {Alexander}}, \bibinfo {author} {\bibfnamefont {M.}~\bibnamefont {Vasmer}},
  \bibinfo {author} {\bibfnamefont {A.}~\bibnamefont {Patil}}, \bibinfo
  {author} {\bibfnamefont {I.}~\bibnamefont {Tzitrin}}, \bibinfo {author}
  {\bibfnamefont {T.}~\bibnamefont {Matsuura}}, \bibinfo {author}
  {\bibfnamefont {D.}~\bibnamefont {Su}}, \bibinfo {author} {\bibfnamefont
  {B.~Q.}\ \bibnamefont {Baragiola}}, \bibinfo {author} {\bibfnamefont
  {S.}~\bibnamefont {Guha}}, \bibinfo {author} {\bibfnamefont {G.}~\bibnamefont
  {Dauphinais}}, \bibinfo {author} {\bibfnamefont {K.~K.}\ \bibnamefont
  {Sabapathy}}, \bibinfo {author} {\bibfnamefont {N.~C.}\ \bibnamefont
  {Menicucci}},\ and\ \bibinfo {author} {\bibfnamefont {I.}~\bibnamefont
  {Dhand}},\ }\bibfield  {title} {\bibinfo {title} {Blueprint for a scalable
  photonic fault-tolerant quantum computer},\ }\href
  {https://doi.org/10.22331/q-2021-02-04-392} {\bibfield  {journal} {\bibinfo
  {journal} {Quantum}\ }\textbf {\bibinfo {volume} {5}},\ \bibinfo {pages}
  {392} (\bibinfo {year} {2021})},\ \Eprint {https://arxiv.org/abs/2010.02905}
  {2010.02905} \BibitemShut {NoStop}%
\bibitem [{\citenamefont {Rozpedek}\ \emph {et~al.}(2021)\citenamefont
  {Rozpedek}, \citenamefont {Noh}, \citenamefont {Xu}, \citenamefont {Guha},\
  and\ \citenamefont {Jiang}}]{rozpedek_quantum_2021}%
  \BibitemOpen
  \bibfield  {author} {\bibinfo {author} {\bibfnamefont {F.}~\bibnamefont
  {Rozpedek}}, \bibinfo {author} {\bibfnamefont {K.}~\bibnamefont {Noh}},
  \bibinfo {author} {\bibfnamefont {Q.}~\bibnamefont {Xu}}, \bibinfo {author}
  {\bibfnamefont {S.}~\bibnamefont {Guha}},\ and\ \bibinfo {author}
  {\bibfnamefont {L.}~\bibnamefont {Jiang}},\ }\bibfield  {title} {\bibinfo
  {title} {Quantum repeaters based on concatenated bosonic and
  discrete-variable quantum codes},\ }\href
  {https://doi.org/10.1038/s41534-021-00438-7} {\bibfield  {journal} {\bibinfo
  {journal} {npj Quantum Information}\ }\textbf {\bibinfo {volume} {7}},\
  \bibinfo {pages} {102} (\bibinfo {year} {2021})}\BibitemShut {NoStop}%
\bibitem [{\citenamefont {Duivenvoorden}\ \emph {et~al.}(2017)\citenamefont
  {Duivenvoorden}, \citenamefont {Terhal},\ and\ \citenamefont
  {Weigand}}]{Duivenvoorden_2017}%
  \BibitemOpen
  \bibfield  {author} {\bibinfo {author} {\bibfnamefont {K.}~\bibnamefont
  {Duivenvoorden}}, \bibinfo {author} {\bibfnamefont {B.~M.}\ \bibnamefont
  {Terhal}},\ and\ \bibinfo {author} {\bibfnamefont {D.}~\bibnamefont
  {Weigand}},\ }\bibfield  {title} {\bibinfo {title} {Single-mode displacement
  sensor},\ }\bibfield  {journal} {\bibinfo  {journal} {Physical Review A}\
  }\textbf {\bibinfo {volume} {95}},\ \href
  {https://doi.org/10.1103/physreva.95.012305} {10.1103/physreva.95.012305}
  (\bibinfo {year} {2017})\BibitemShut {NoStop}%
\bibitem [{\citenamefont {Terhal}\ and\ \citenamefont
  {Weigand}(2016)}]{Terhal_2016}%
  \BibitemOpen
  \bibfield  {author} {\bibinfo {author} {\bibfnamefont {B.~M.}\ \bibnamefont
  {Terhal}}\ and\ \bibinfo {author} {\bibfnamefont {D.}~\bibnamefont
  {Weigand}},\ }\bibfield  {title} {\bibinfo {title} {Encoding a qubit into a
  cavity mode in circuit {QED} using phase estimation},\ }\bibfield  {journal}
  {\bibinfo  {journal} {Physical Review A}\ }\textbf {\bibinfo {volume} {93}},\
  \href {https://doi.org/10.1103/physreva.93.012315}
  {10.1103/physreva.93.012315} (\bibinfo {year} {2016})\BibitemShut {NoStop}%
\bibitem [{\citenamefont {Fabre}\ \emph
  {et~al.}(2020{\natexlab{b}})\citenamefont {Fabre}, \citenamefont {Belhassen},
  \citenamefont {Minneci}, \citenamefont {Felicetti}, \citenamefont {Keller},
  \citenamefont {Amanti}, \citenamefont {Baboux}, \citenamefont {Coudreau},
  \citenamefont {Ducci},\ and\ \citenamefont {Milman}}]{PhysRevA.102.023710}%
  \BibitemOpen
  \bibfield  {author} {\bibinfo {author} {\bibfnamefont {N.}~\bibnamefont
  {Fabre}}, \bibinfo {author} {\bibfnamefont {J.}~\bibnamefont {Belhassen}},
  \bibinfo {author} {\bibfnamefont {A.}~\bibnamefont {Minneci}}, \bibinfo
  {author} {\bibfnamefont {S.}~\bibnamefont {Felicetti}}, \bibinfo {author}
  {\bibfnamefont {A.}~\bibnamefont {Keller}}, \bibinfo {author} {\bibfnamefont
  {M.~I.}\ \bibnamefont {Amanti}}, \bibinfo {author} {\bibfnamefont
  {F.}~\bibnamefont {Baboux}}, \bibinfo {author} {\bibfnamefont
  {T.}~\bibnamefont {Coudreau}}, \bibinfo {author} {\bibfnamefont
  {S.}~\bibnamefont {Ducci}},\ and\ \bibinfo {author} {\bibfnamefont
  {P.}~\bibnamefont {Milman}},\ }\bibfield  {title} {\bibinfo {title}
  {Producing a delocalized frequency-time schr\"odinger-cat-like state with
  hong-ou-mandel interferometry},\ }\href
  {https://doi.org/10.1103/PhysRevA.102.023710} {\bibfield  {journal} {\bibinfo
   {journal} {Phys. Rev. A}\ }\textbf {\bibinfo {volume} {102}},\ \bibinfo
  {pages} {023710} (\bibinfo {year} {2020}{\natexlab{b}})}\BibitemShut
  {NoStop}%
\bibitem [{\citenamefont {Yamazaki}\ \emph {et~al.}(2023)\citenamefont
  {Yamazaki}, \citenamefont {Arizono}, \citenamefont {Kobayashi}, \citenamefont
  {Ikuta},\ and\ \citenamefont
  {Yamamoto}}]{https://doi.org/10.48550/arxiv.2301.03188}%
  \BibitemOpen
  \bibfield  {author} {\bibinfo {author} {\bibfnamefont {T.}~\bibnamefont
  {Yamazaki}}, \bibinfo {author} {\bibfnamefont {T.}~\bibnamefont {Arizono}},
  \bibinfo {author} {\bibfnamefont {T.}~\bibnamefont {Kobayashi}}, \bibinfo
  {author} {\bibfnamefont {R.}~\bibnamefont {Ikuta}},\ and\ \bibinfo {author}
  {\bibfnamefont {T.}~\bibnamefont {Yamamoto}},\ }\href
  {https://doi.org/10.48550/ARXIV.2301.03188} {\bibinfo {title} {Linear optical
  quantum computation with frequency-comb qubits and passive devices}}
  (\bibinfo {year} {2023})\BibitemShut {NoStop}%
\bibitem [{\citenamefont {Walshe}\ \emph {et~al.}(2020)\citenamefont {Walshe},
  \citenamefont {Baragiola}, \citenamefont {Alexander},\ and\ \citenamefont
  {Menicucci}}]{PhysRevA.102.062411}%
  \BibitemOpen
  \bibfield  {author} {\bibinfo {author} {\bibfnamefont {B.~W.}\ \bibnamefont
  {Walshe}}, \bibinfo {author} {\bibfnamefont {B.~Q.}\ \bibnamefont
  {Baragiola}}, \bibinfo {author} {\bibfnamefont {R.~N.}\ \bibnamefont
  {Alexander}},\ and\ \bibinfo {author} {\bibfnamefont {N.~C.}\ \bibnamefont
  {Menicucci}},\ }\bibfield  {title} {\bibinfo {title} {Continuous-variable
  gate teleportation and bosonic-code error correction},\ }\href
  {https://doi.org/10.1103/PhysRevA.102.062411} {\bibfield  {journal} {\bibinfo
   {journal} {Phys. Rev. A}\ }\textbf {\bibinfo {volume} {102}},\ \bibinfo
  {pages} {062411} (\bibinfo {year} {2020})}\BibitemShut {NoStop}%
\bibitem [{\citenamefont {Fukui}\ \emph {et~al.}(2021)\citenamefont {Fukui},
  \citenamefont {Alexander},\ and\ \citenamefont {van
  Loock}}]{PhysRevResearch.3.033118}%
  \BibitemOpen
  \bibfield  {author} {\bibinfo {author} {\bibfnamefont {K.}~\bibnamefont
  {Fukui}}, \bibinfo {author} {\bibfnamefont {R.~N.}\ \bibnamefont
  {Alexander}},\ and\ \bibinfo {author} {\bibfnamefont {P.}~\bibnamefont {van
  Loock}},\ }\bibfield  {title} {\bibinfo {title} {All-optical long-distance
  quantum communication with gottesman-kitaev-preskill qubits},\ }\href
  {https://doi.org/10.1103/PhysRevResearch.3.033118} {\bibfield  {journal}
  {\bibinfo  {journal} {Phys. Rev. Research}\ }\textbf {\bibinfo {volume}
  {3}},\ \bibinfo {pages} {033118} (\bibinfo {year} {2021})}\BibitemShut
  {NoStop}%
\bibitem [{\citenamefont {Sisodia}\ \emph {et~al.}(2017)\citenamefont
  {Sisodia}, \citenamefont {Verma}, \citenamefont {Thapliyal},\ and\
  \citenamefont {Pathak}}]{sisodia_teleportation_2017}%
  \BibitemOpen
  \bibfield  {author} {\bibinfo {author} {\bibfnamefont {M.}~\bibnamefont
  {Sisodia}}, \bibinfo {author} {\bibfnamefont {V.}~\bibnamefont {Verma}},
  \bibinfo {author} {\bibfnamefont {K.}~\bibnamefont {Thapliyal}},\ and\
  \bibinfo {author} {\bibfnamefont {A.}~\bibnamefont {Pathak}},\ }\bibfield
  {title} {\bibinfo {title} {Teleportation of a qubit using entangled
  non-orthogonal states: a comparative study},\ }\href
  {https://doi.org/10.1007/s11128-017-1526-x} {\bibfield  {journal} {\bibinfo
  {journal} {Quantum Inf Process}\ }\textbf {\bibinfo {volume} {16}},\ \bibinfo
  {pages} {76} (\bibinfo {year} {2017})}\BibitemShut {NoStop}%
\bibitem [{\citenamefont {Bennett}\ \emph {et~al.}(1993)\citenamefont
  {Bennett}, \citenamefont {Brassard}, \citenamefont {Cr\'epeau}, \citenamefont
  {Jozsa}, \citenamefont {Peres},\ and\ \citenamefont
  {Wootters}}]{PhysRevLett.70.1895}%
  \BibitemOpen
  \bibfield  {author} {\bibinfo {author} {\bibfnamefont {C.~H.}\ \bibnamefont
  {Bennett}}, \bibinfo {author} {\bibfnamefont {G.}~\bibnamefont {Brassard}},
  \bibinfo {author} {\bibfnamefont {C.}~\bibnamefont {Cr\'epeau}}, \bibinfo
  {author} {\bibfnamefont {R.}~\bibnamefont {Jozsa}}, \bibinfo {author}
  {\bibfnamefont {A.}~\bibnamefont {Peres}},\ and\ \bibinfo {author}
  {\bibfnamefont {W.~K.}\ \bibnamefont {Wootters}},\ }\bibfield  {title}
  {\bibinfo {title} {Teleporting an unknown quantum state via dual classical
  and einstein-podolsky-rosen channels},\ }\href
  {https://doi.org/10.1103/PhysRevLett.70.1895} {\bibfield  {journal} {\bibinfo
   {journal} {Phys. Rev. Lett.}\ }\textbf {\bibinfo {volume} {70}},\ \bibinfo
  {pages} {1895} (\bibinfo {year} {1993})}\BibitemShut {NoStop}%
\bibitem [{\citenamefont {Bouwmeester}\ \emph {et~al.}(2000)\citenamefont
  {Bouwmeester}, \citenamefont {Pan}, \citenamefont {Weinfurter},\ and\
  \citenamefont {Zeilinger}}]{bouwmeester_high-fidelity_2000}%
  \BibitemOpen
  \bibfield  {author} {\bibinfo {author} {\bibfnamefont {D.}~\bibnamefont
  {Bouwmeester}}, \bibinfo {author} {\bibfnamefont {J.-W.}\ \bibnamefont
  {Pan}}, \bibinfo {author} {\bibfnamefont {H.}~\bibnamefont {Weinfurter}},\
  and\ \bibinfo {author} {\bibfnamefont {A.}~\bibnamefont {Zeilinger}},\
  }\bibfield  {title} {\bibinfo {title} {High-fidelity teleportation of
  independent qubits},\ }\href {https://doi.org/10.1080/09500340008244042}
  {\bibfield  {journal} {\bibinfo  {journal} {Journal of Modern Optics}\
  }\textbf {\bibinfo {volume} {47}},\ \bibinfo {pages} {279} (\bibinfo {year}
  {2000})},\ \Eprint {https://arxiv.org/abs/quant-ph/9910043}
  {quant-ph/9910043} \BibitemShut {NoStop}%
\bibitem [{\citenamefont {Zhong}\ \emph {et~al.}(2015)\citenamefont {Zhong},
  \citenamefont {Zhou}, \citenamefont {Horansky}, \citenamefont {Lee},
  \citenamefont {Verma}, \citenamefont {Lita}, \citenamefont {Restelli},
  \citenamefont {Bienfang}, \citenamefont {Mirin}, \citenamefont {Gerrits},
  \citenamefont {Nam}, \citenamefont {Marsili}, \citenamefont {Shaw},
  \citenamefont {Zhang}, \citenamefont {Wang}, \citenamefont {Englund},
  \citenamefont {Wornell}, \citenamefont {Shapiro},\ and\ \citenamefont
  {Wong}}]{Zhong_2015}%
  \BibitemOpen
  \bibfield  {author} {\bibinfo {author} {\bibfnamefont {T.}~\bibnamefont
  {Zhong}}, \bibinfo {author} {\bibfnamefont {H.}~\bibnamefont {Zhou}},
  \bibinfo {author} {\bibfnamefont {R.~D.}\ \bibnamefont {Horansky}}, \bibinfo
  {author} {\bibfnamefont {C.}~\bibnamefont {Lee}}, \bibinfo {author}
  {\bibfnamefont {V.~B.}\ \bibnamefont {Verma}}, \bibinfo {author}
  {\bibfnamefont {A.~E.}\ \bibnamefont {Lita}}, \bibinfo {author}
  {\bibfnamefont {A.}~\bibnamefont {Restelli}}, \bibinfo {author}
  {\bibfnamefont {J.~C.}\ \bibnamefont {Bienfang}}, \bibinfo {author}
  {\bibfnamefont {R.~P.}\ \bibnamefont {Mirin}}, \bibinfo {author}
  {\bibfnamefont {T.}~\bibnamefont {Gerrits}}, \bibinfo {author} {\bibfnamefont
  {S.~W.}\ \bibnamefont {Nam}}, \bibinfo {author} {\bibfnamefont
  {F.}~\bibnamefont {Marsili}}, \bibinfo {author} {\bibfnamefont {M.~D.}\
  \bibnamefont {Shaw}}, \bibinfo {author} {\bibfnamefont {Z.}~\bibnamefont
  {Zhang}}, \bibinfo {author} {\bibfnamefont {L.}~\bibnamefont {Wang}},
  \bibinfo {author} {\bibfnamefont {D.}~\bibnamefont {Englund}}, \bibinfo
  {author} {\bibfnamefont {G.~W.}\ \bibnamefont {Wornell}}, \bibinfo {author}
  {\bibfnamefont {J.~H.}\ \bibnamefont {Shapiro}},\ and\ \bibinfo {author}
  {\bibfnamefont {F.~N.~C.}\ \bibnamefont {Wong}},\ }\bibfield  {title}
  {\bibinfo {title} {Photon-efficient quantum key distribution using
  time–energy entanglement with high-dimensional encoding},\ }\href
  {https://doi.org/10.1088/1367-2630/17/2/022002} {\bibfield  {journal}
  {\bibinfo  {journal} {New Journal of Physics}\ }\textbf {\bibinfo {volume}
  {17}},\ \bibinfo {pages} {022002} (\bibinfo {year} {2015})}\BibitemShut
  {NoStop}%
\bibitem [{\citenamefont {Jin}\ \emph {et~al.}(2019)\citenamefont {Jin},
  \citenamefont {Bourgoin}, \citenamefont {Tannous}, \citenamefont {Agne},
  \citenamefont {Pugh}, \citenamefont {Kuntz}, \citenamefont {Higgins},\ and\
  \citenamefont {Jennewein}}]{Jin:19}%
  \BibitemOpen
  \bibfield  {author} {\bibinfo {author} {\bibfnamefont {J.}~\bibnamefont
  {Jin}}, \bibinfo {author} {\bibfnamefont {J.-P.}\ \bibnamefont {Bourgoin}},
  \bibinfo {author} {\bibfnamefont {R.}~\bibnamefont {Tannous}}, \bibinfo
  {author} {\bibfnamefont {S.}~\bibnamefont {Agne}}, \bibinfo {author}
  {\bibfnamefont {C.~J.}\ \bibnamefont {Pugh}}, \bibinfo {author}
  {\bibfnamefont {K.~B.}\ \bibnamefont {Kuntz}}, \bibinfo {author}
  {\bibfnamefont {B.~L.}\ \bibnamefont {Higgins}},\ and\ \bibinfo {author}
  {\bibfnamefont {T.}~\bibnamefont {Jennewein}},\ }\bibfield  {title} {\bibinfo
  {title} {Genuine time-bin-encoded quantum key distribution over a turbulent
  depolarizing free-space channel},\ }\href
  {https://doi.org/10.1364/OE.27.037214} {\bibfield  {journal} {\bibinfo
  {journal} {Opt. Express}\ }\textbf {\bibinfo {volume} {27}},\ \bibinfo
  {pages} {37214} (\bibinfo {year} {2019})}\BibitemShut {NoStop}%
\bibitem [{\citenamefont {Vagniluca}\ \emph {et~al.}(2020)\citenamefont
  {Vagniluca}, \citenamefont {Da~Lio}, \citenamefont {Rusca}, \citenamefont
  {Cozzolino}, \citenamefont {Ding}, \citenamefont {Zbinden}, \citenamefont
  {Zavatta}, \citenamefont {Oxenl\o{}we},\ and\ \citenamefont
  {Bacco}}]{PhysRevApplied.14.014051}%
  \BibitemOpen
  \bibfield  {author} {\bibinfo {author} {\bibfnamefont {I.}~\bibnamefont
  {Vagniluca}}, \bibinfo {author} {\bibfnamefont {B.}~\bibnamefont {Da~Lio}},
  \bibinfo {author} {\bibfnamefont {D.}~\bibnamefont {Rusca}}, \bibinfo
  {author} {\bibfnamefont {D.}~\bibnamefont {Cozzolino}}, \bibinfo {author}
  {\bibfnamefont {Y.}~\bibnamefont {Ding}}, \bibinfo {author} {\bibfnamefont
  {H.}~\bibnamefont {Zbinden}}, \bibinfo {author} {\bibfnamefont
  {A.}~\bibnamefont {Zavatta}}, \bibinfo {author} {\bibfnamefont {L.~K.}\
  \bibnamefont {Oxenl\o{}we}},\ and\ \bibinfo {author} {\bibfnamefont
  {D.}~\bibnamefont {Bacco}},\ }\bibfield  {title} {\bibinfo {title} {Efficient
  time-bin encoding for practical high-dimensional quantum key distribution},\
  }\href {https://doi.org/10.1103/PhysRevApplied.14.014051} {\bibfield
  {journal} {\bibinfo  {journal} {Phys. Rev. Appl.}\ }\textbf {\bibinfo
  {volume} {14}},\ \bibinfo {pages} {014051} (\bibinfo {year}
  {2020})}\BibitemShut {NoStop}%
\bibitem [{\citenamefont {Lukens}\ and\ \citenamefont
  {Lougovski}(2017)}]{lukens_frequency-encoded_2017}%
  \BibitemOpen
  \bibfield  {author} {\bibinfo {author} {\bibfnamefont {J.~M.}\ \bibnamefont
  {Lukens}}\ and\ \bibinfo {author} {\bibfnamefont {P.}~\bibnamefont
  {Lougovski}},\ }\bibfield  {title} {\bibinfo {title} {Frequency-encoded
  photonic qubits for scalable quantum information processing},\ }\href
  {https://doi.org/10.1364/OPTICA.4.000008} {\bibfield  {journal} {\bibinfo
  {journal} {Optica}\ }\textbf {\bibinfo {volume} {4}},\ \bibinfo {pages} {8}
  (\bibinfo {year} {2017})}\BibitemShut {NoStop}%
\bibitem [{\citenamefont {Lu}\ \emph {et~al.}(2019)\citenamefont {Lu},
  \citenamefont {Lukens}, \citenamefont {Williams}, \citenamefont {Imany},
  \citenamefont {Peters}, \citenamefont {Weiner},\ and\ \citenamefont
  {Lougovski}}]{lu_controlled-not_2019}%
  \BibitemOpen
  \bibfield  {author} {\bibinfo {author} {\bibfnamefont {H.-H.}\ \bibnamefont
  {Lu}}, \bibinfo {author} {\bibfnamefont {J.~M.}\ \bibnamefont {Lukens}},
  \bibinfo {author} {\bibfnamefont {B.~P.}\ \bibnamefont {Williams}}, \bibinfo
  {author} {\bibfnamefont {P.}~\bibnamefont {Imany}}, \bibinfo {author}
  {\bibfnamefont {N.~A.}\ \bibnamefont {Peters}}, \bibinfo {author}
  {\bibfnamefont {A.~M.}\ \bibnamefont {Weiner}},\ and\ \bibinfo {author}
  {\bibfnamefont {P.}~\bibnamefont {Lougovski}},\ }\bibfield  {title} {\bibinfo
  {title} {A controlled-{NOT} gate for frequency-bin qubits},\ }\href
  {https://doi.org/10.1038/s41534-019-0137-z} {\bibfield  {journal} {\bibinfo
  {journal} {npj Quantum Information}\ }\textbf {\bibinfo {volume} {5}},\
  \bibinfo {pages} {24} (\bibinfo {year} {2019})}\BibitemShut {NoStop}%
\bibitem [{\citenamefont {Francesconi}\ \emph {et~al.}(2022)\citenamefont
  {Francesconi}, \citenamefont {Raymond}, \citenamefont {Duhamel},
  \citenamefont {Filloux}, \citenamefont {Lemaître}, \citenamefont {Milman},
  \citenamefont {Amanti}, \citenamefont {Baboux},\ and\ \citenamefont
  {Ducci}}]{https://doi.org/10.48550/arxiv.2207.10943}%
  \BibitemOpen
  \bibfield  {author} {\bibinfo {author} {\bibfnamefont {S.}~\bibnamefont
  {Francesconi}}, \bibinfo {author} {\bibfnamefont {A.}~\bibnamefont
  {Raymond}}, \bibinfo {author} {\bibfnamefont {R.}~\bibnamefont {Duhamel}},
  \bibinfo {author} {\bibfnamefont {P.}~\bibnamefont {Filloux}}, \bibinfo
  {author} {\bibfnamefont {A.}~\bibnamefont {Lemaître}}, \bibinfo {author}
  {\bibfnamefont {P.}~\bibnamefont {Milman}}, \bibinfo {author} {\bibfnamefont
  {M.~I.}\ \bibnamefont {Amanti}}, \bibinfo {author} {\bibfnamefont
  {F.}~\bibnamefont {Baboux}},\ and\ \bibinfo {author} {\bibfnamefont
  {S.}~\bibnamefont {Ducci}},\ }\href
  {https://doi.org/10.48550/ARXIV.2207.10943} {\bibinfo {title} {On-chip
  generation of hybrid polarization-frequency entangled biphoton states}}
  (\bibinfo {year} {2022})\BibitemShut {NoStop}%
\bibitem [{\citenamefont {Chen}\ \emph {et~al.}(2019)\citenamefont {Chen},
  \citenamefont {Fink}, \citenamefont {Steinlechner}, \citenamefont {Torres},\
  and\ \citenamefont {Ursin}}]{chen_hong-ou-mandel_2019}%
  \BibitemOpen
  \bibfield  {author} {\bibinfo {author} {\bibfnamefont {Y.}~\bibnamefont
  {Chen}}, \bibinfo {author} {\bibfnamefont {M.}~\bibnamefont {Fink}}, \bibinfo
  {author} {\bibfnamefont {F.}~\bibnamefont {Steinlechner}}, \bibinfo {author}
  {\bibfnamefont {J.~P.}\ \bibnamefont {Torres}},\ and\ \bibinfo {author}
  {\bibfnamefont {R.}~\bibnamefont {Ursin}},\ }\bibfield  {title} {\bibinfo
  {title} {Hong-{Ou}-{Mandel} interferometry on a biphoton beat note},\ }\href
  {https://doi.org/10.1038/s41534-019-0161-z} {\bibfield  {journal} {\bibinfo
  {journal} {npj Quantum Information}\ }\textbf {\bibinfo {volume} {5}},\
  \bibinfo {pages} {43} (\bibinfo {year} {2019})}\BibitemShut {NoStop}%
\bibitem [{\citenamefont {Jayakumar}\ \emph {et~al.}(2014)\citenamefont
  {Jayakumar}, \citenamefont {Predojević}, \citenamefont {Kauten},
  \citenamefont {Huber}, \citenamefont {Solomon},\ and\ \citenamefont
  {Weihs}}]{jayakumar_time-bin_2014}%
  \BibitemOpen
  \bibfield  {author} {\bibinfo {author} {\bibfnamefont {H.}~\bibnamefont
  {Jayakumar}}, \bibinfo {author} {\bibfnamefont {A.}~\bibnamefont
  {Predojević}}, \bibinfo {author} {\bibfnamefont {T.}~\bibnamefont {Kauten}},
  \bibinfo {author} {\bibfnamefont {T.}~\bibnamefont {Huber}}, \bibinfo
  {author} {\bibfnamefont {G.~S.}\ \bibnamefont {Solomon}},\ and\ \bibinfo
  {author} {\bibfnamefont {G.}~\bibnamefont {Weihs}},\ }\bibfield  {title}
  {\bibinfo {title} {Time-bin entangled photons from a quantum dot},\ }\href
  {https://doi.org/10.1038/ncomms5251} {\bibfield  {journal} {\bibinfo
  {journal} {Nature Communications}\ }\textbf {\bibinfo {volume} {5}},\
  \bibinfo {pages} {4251} (\bibinfo {year} {2014})}\BibitemShut {NoStop}%
\bibitem [{\citenamefont {Kim}\ \emph {et~al.}(2022)\citenamefont {Kim},
  \citenamefont {Chae}, \citenamefont {Jeong},\ and\ \citenamefont
  {Kim}}]{kim_quantum_2022}%
  \BibitemOpen
  \bibfield  {author} {\bibinfo {author} {\bibfnamefont {J.-H.}\ \bibnamefont
  {Kim}}, \bibinfo {author} {\bibfnamefont {J.-W.}\ \bibnamefont {Chae}},
  \bibinfo {author} {\bibfnamefont {Y.-C.}\ \bibnamefont {Jeong}},\ and\
  \bibinfo {author} {\bibfnamefont {Y.-H.}\ \bibnamefont {Kim}},\ }\bibfield
  {title} {\bibinfo {title} {Quantum communication with time-bin entanglement
  over a wavelength-multiplexed fiber network},\ }\href
  {https://doi.org/10.1063/5.0073040} {\bibfield  {journal} {\bibinfo
  {journal} {APL Photonics}\ }\textbf {\bibinfo {volume} {7}},\ \bibinfo
  {pages} {016106} (\bibinfo {year} {2022})},\ \Eprint
  {https://arxiv.org/abs/https://doi.org/10.1063/5.0073040}
  {https://doi.org/10.1063/5.0073040} \BibitemShut {NoStop}%
\bibitem [{\citenamefont {Ketterer}\ \emph {et~al.}(2016)\citenamefont
  {Ketterer}, \citenamefont {Keller}, \citenamefont {Walborn}, \citenamefont
  {Coudreau},\ and\ \citenamefont {Milman}}]{PhysRevA.94.022325}%
  \BibitemOpen
  \bibfield  {author} {\bibinfo {author} {\bibfnamefont {A.}~\bibnamefont
  {Ketterer}}, \bibinfo {author} {\bibfnamefont {A.}~\bibnamefont {Keller}},
  \bibinfo {author} {\bibfnamefont {S.~P.}\ \bibnamefont {Walborn}}, \bibinfo
  {author} {\bibfnamefont {T.}~\bibnamefont {Coudreau}},\ and\ \bibinfo
  {author} {\bibfnamefont {P.}~\bibnamefont {Milman}},\ }\bibfield  {title}
  {\bibinfo {title} {Quantum information processing in phase space: A modular
  variables approach},\ }\href {https://doi.org/10.1103/PhysRevA.94.022325}
  {\bibfield  {journal} {\bibinfo  {journal} {Phys. Rev. A}\ }\textbf {\bibinfo
  {volume} {94}},\ \bibinfo {pages} {022325} (\bibinfo {year}
  {2016})}\BibitemShut {NoStop}%
\bibitem [{\citenamefont {Albert}\ \emph {et~al.}(2018)\citenamefont {Albert},
  \citenamefont {Noh}, \citenamefont {Duivenvoorden}, \citenamefont {Young},
  \citenamefont {Brierley}, \citenamefont {Reinhold}, \citenamefont {Vuillot},
  \citenamefont {Li}, \citenamefont {Shen}, \citenamefont {Girvin},
  \citenamefont {Terhal},\ and\ \citenamefont
  {Jiang}}]{albert_performance_2018}%
  \BibitemOpen
  \bibfield  {author} {\bibinfo {author} {\bibfnamefont {V.~V.}\ \bibnamefont
  {Albert}}, \bibinfo {author} {\bibfnamefont {K.}~\bibnamefont {Noh}},
  \bibinfo {author} {\bibfnamefont {K.}~\bibnamefont {Duivenvoorden}}, \bibinfo
  {author} {\bibfnamefont {D.~J.}\ \bibnamefont {Young}}, \bibinfo {author}
  {\bibfnamefont {R.~T.}\ \bibnamefont {Brierley}}, \bibinfo {author}
  {\bibfnamefont {P.}~\bibnamefont {Reinhold}}, \bibinfo {author}
  {\bibfnamefont {C.}~\bibnamefont {Vuillot}}, \bibinfo {author} {\bibfnamefont
  {L.}~\bibnamefont {Li}}, \bibinfo {author} {\bibfnamefont {C.}~\bibnamefont
  {Shen}}, \bibinfo {author} {\bibfnamefont {S.~M.}\ \bibnamefont {Girvin}},
  \bibinfo {author} {\bibfnamefont {B.~M.}\ \bibnamefont {Terhal}},\ and\
  \bibinfo {author} {\bibfnamefont {L.}~\bibnamefont {Jiang}},\ }\bibfield
  {title} {\bibinfo {title} {Performance and structure of single-mode bosonic
  codes},\ }\href {https://doi.org/10.1103/PhysRevA.97.032346} {\bibfield
  {journal} {\bibinfo  {journal} {Phys. Rev. A}\ }\textbf {\bibinfo {volume}
  {97}},\ \bibinfo {pages} {032346} (\bibinfo {year} {2018})},\ \Eprint
  {https://arxiv.org/abs/1708.05010} {1708.05010} \BibitemShut {NoStop}%
\bibitem [{\citenamefont {Hong}\ \emph {et~al.}(2018)\citenamefont {Hong},
  \citenamefont {Baek}, \citenamefont {Kwon},\ and\ \citenamefont
  {Kim}}]{hong_dispersive_2018}%
  \BibitemOpen
  \bibfield  {author} {\bibinfo {author} {\bibfnamefont {K.-H.}\ \bibnamefont
  {Hong}}, \bibinfo {author} {\bibfnamefont {S.-Y.}\ \bibnamefont {Baek}},
  \bibinfo {author} {\bibfnamefont {O.}~\bibnamefont {Kwon}},\ and\ \bibinfo
  {author} {\bibfnamefont {Y.-H.}\ \bibnamefont {Kim}},\ }\bibfield  {title}
  {\bibinfo {title} {Dispersive broadening of two-photon wave packets generated
  via type-i and type-{II} spontaneous parametric down-conversion},\ }\href
  {https://doi.org/10.3938/jkps.73.1650} {\bibfield  {journal} {\bibinfo
  {journal} {J. Korean Phys. Soc.}\ }\textbf {\bibinfo {volume} {73}},\
  \bibinfo {pages} {1650} (\bibinfo {year} {2018})}\BibitemShut {NoStop}%
\bibitem [{\citenamefont {Maram}\ and\ \citenamefont
  {Azaña}(2013)}]{maram_spectral_2013}%
  \BibitemOpen
  \bibfield  {author} {\bibinfo {author} {\bibfnamefont {R.}~\bibnamefont
  {Maram}}\ and\ \bibinfo {author} {\bibfnamefont {J.}~\bibnamefont {Azaña}},\
  }\bibfield  {title} {\bibinfo {title} {Spectral self-imaging of time-periodic
  coherent frequency combs by parabolic cross-phase modulation},\ }\href
  {https://doi.org/10.1364/OE.21.028824} {\bibfield  {journal} {\bibinfo
  {journal} {Opt. Express}\ }\textbf {\bibinfo {volume} {21}},\ \bibinfo
  {pages} {28824} (\bibinfo {year} {2013})}\BibitemShut {NoStop}%
\bibitem [{\citenamefont {Antonelli}\ and\ \citenamefont
  {Mecozzi}(2005)}]{antonelli_pulse_2005}%
  \BibitemOpen
  \bibfield  {author} {\bibinfo {author} {\bibfnamefont {C.}~\bibnamefont
  {Antonelli}}\ and\ \bibinfo {author} {\bibfnamefont {A.}~\bibnamefont
  {Mecozzi}},\ }\bibfield  {title} {\bibinfo {title} {Pulse broadening due to
  polarization mode dispersion with first-order compensation},\ }\href
  {https://doi.org/10.1364/OL.30.001626} {\bibfield  {journal} {\bibinfo
  {journal} {Opt. Lett.}\ }\textbf {\bibinfo {volume} {30}},\ \bibinfo {pages}
  {1626} (\bibinfo {year} {2005})}\BibitemShut {NoStop}%
\bibitem [{\citenamefont {Poon}\ and\ \citenamefont
  {Law}(2008)}]{poon_polarization_2008}%
  \BibitemOpen
  \bibfield  {author} {\bibinfo {author} {\bibfnamefont {P.~S.~Y.}\
  \bibnamefont {Poon}}\ and\ \bibinfo {author} {\bibfnamefont {C.~K.}\
  \bibnamefont {Law}},\ }\bibfield  {title} {\bibinfo {title} {Polarization and
  frequency disentanglement of photons via stochastic polarization mode
  dispersion},\ }\href {https://doi.org/10.1103/PhysRevA.77.032330} {\bibfield
  {journal} {\bibinfo  {journal} {Phys. Rev. A}\ }\textbf {\bibinfo {volume}
  {77}},\ \bibinfo {pages} {032330} (\bibinfo {year} {2008})}\BibitemShut
  {NoStop}%
\bibitem [{\citenamefont {Gordon}\ and\ \citenamefont
  {Kogelnik}(2000)}]{gordon_pmd_2000}%
  \BibitemOpen
  \bibfield  {author} {\bibinfo {author} {\bibfnamefont {J.~P.}\ \bibnamefont
  {Gordon}}\ and\ \bibinfo {author} {\bibfnamefont {H.}~\bibnamefont
  {Kogelnik}},\ }\bibfield  {title} {\bibinfo {title} {{PMD} fundamentals:
  Polarization mode dispersion in optical fibers},\ }\href
  {https://doi.org/10.1073/pnas.97.9.4541} {\bibfield  {journal} {\bibinfo
  {journal} {Proceedings of the National Academy of Sciences}\ }\textbf
  {\bibinfo {volume} {97}},\ \bibinfo {pages} {4541} (\bibinfo {year}
  {2000})}\BibitemShut {NoStop}%
\bibitem [{\citenamefont {Chang-hua}\ \emph {et~al.}(2009)\citenamefont
  {Chang-hua}, \citenamefont {Chang-xing}, \citenamefont {Dong-xiao},
  \citenamefont {Nan},\ and\ \citenamefont
  {Yun-hui}}]{chang-hua_polarization_nodate}%
  \BibitemOpen
  \bibfield  {author} {\bibinfo {author} {\bibfnamefont {Z.}~\bibnamefont
  {Chang-hua}}, \bibinfo {author} {\bibfnamefont {P.}~\bibnamefont
  {Chang-xing}}, \bibinfo {author} {\bibfnamefont {Q.}~\bibnamefont
  {Dong-xiao}}, \bibinfo {author} {\bibfnamefont {C.}~\bibnamefont {Nan}},\
  and\ \bibinfo {author} {\bibfnamefont {Y.}~\bibnamefont {Yun-hui}},\
  }\bibfield  {title} {\bibinfo {title} {Polarization state dynamics of single
  photon pulse},\ }\href@noop {} {\bibfield  {journal} {\bibinfo  {journal}
  {arXiv:0908.4370}\ ,\ \bibinfo {pages} {7}} (\bibinfo {year}
  {2009})}\BibitemShut {NoStop}%
\bibitem [{\citenamefont {Matsuda}(2016)}]{matsuda_deterministic_2016}%
  \BibitemOpen
  \bibfield  {author} {\bibinfo {author} {\bibfnamefont {N.}~\bibnamefont
  {Matsuda}},\ }\bibfield  {title} {\bibinfo {title} {Deterministic reshaping
  of single-photon spectra using cross-phase modulation},\ }\href
  {https://doi.org/10.1126/sciadv.1501223} {\bibfield  {journal} {\bibinfo
  {journal} {Sci. Adv.}\ }\textbf {\bibinfo {volume} {2}},\ \bibinfo {pages}
  {e1501223} (\bibinfo {year} {2016})}\BibitemShut {NoStop}%
\bibitem [{\citenamefont {Kurzyna}\ \emph {et~al.}(2022)\citenamefont
  {Kurzyna}, \citenamefont {Jastrzebski}, \citenamefont {Fabre}, \citenamefont
  {Wasilewski}, \citenamefont {Lipka},\ and\ \citenamefont
  {Parniak}}]{Kurzyna_2022}%
  \BibitemOpen
  \bibfield  {author} {\bibinfo {author} {\bibfnamefont {S.}~\bibnamefont
  {Kurzyna}}, \bibinfo {author} {\bibfnamefont {M.}~\bibnamefont
  {Jastrzebski}}, \bibinfo {author} {\bibfnamefont {N.}~\bibnamefont {Fabre}},
  \bibinfo {author} {\bibfnamefont {W.}~\bibnamefont {Wasilewski}}, \bibinfo
  {author} {\bibfnamefont {M.}~\bibnamefont {Lipka}},\ and\ \bibinfo {author}
  {\bibfnamefont {M.}~\bibnamefont {Parniak}},\ }\bibfield  {title} {\bibinfo
  {title} {Variable electro-optic shearing interferometry for ultrafast
  single-photon-level pulse characterization},\ }\href
  {https://doi.org/10.1364/oe.471108} {\bibfield  {journal} {\bibinfo
  {journal} {Optics Express}\ }\textbf {\bibinfo {volume} {30}},\ \bibinfo
  {pages} {39826} (\bibinfo {year} {2022})}\BibitemShut {NoStop}%
\bibitem [{\citenamefont {Golestani}\ \emph {et~al.}(2022)\citenamefont
  {Golestani}, \citenamefont {Davis}, \citenamefont {So\ifmmode~\acute{s}\else
  \'{s}\fi{}nicki}, \citenamefont {Miko\l{}ajczyk}, \citenamefont {Treps},\
  and\ \citenamefont {Karpi\ifmmode~\acute{n}\else
  \'{n}\fi{}ski}}]{PhysRevLett.129.123605}%
  \BibitemOpen
  \bibfield  {author} {\bibinfo {author} {\bibfnamefont {A.}~\bibnamefont
  {Golestani}}, \bibinfo {author} {\bibfnamefont {A.~O.~C.}\ \bibnamefont
  {Davis}}, \bibinfo {author} {\bibfnamefont {F.}~\bibnamefont
  {So\ifmmode~\acute{s}\else \'{s}\fi{}nicki}}, \bibinfo {author}
  {\bibfnamefont {M.}~\bibnamefont {Miko\l{}ajczyk}}, \bibinfo {author}
  {\bibfnamefont {N.}~\bibnamefont {Treps}},\ and\ \bibinfo {author}
  {\bibfnamefont {M.}~\bibnamefont {Karpi\ifmmode~\acute{n}\else
  \'{n}\fi{}ski}},\ }\bibfield  {title} {\bibinfo {title} {Electro-optic
  fourier transform chronometry of pulsed quantum light},\ }\href
  {https://doi.org/10.1103/PhysRevLett.129.123605} {\bibfield  {journal}
  {\bibinfo  {journal} {Phys. Rev. Lett.}\ }\textbf {\bibinfo {volume} {129}},\
  \bibinfo {pages} {123605} (\bibinfo {year} {2022})}\BibitemShut {NoStop}%
\bibitem [{\citenamefont {Amari}\ \emph {et~al.}(2010)\citenamefont {Amari},
  \citenamefont {Walther}, \citenamefont {Sabooni}, \citenamefont {Huang},
  \citenamefont {Kröll}, \citenamefont {Afzelius}, \citenamefont {Usmani},
  \citenamefont {Lauritzen}, \citenamefont {Sangouard}, \citenamefont
  {de~Riedmatten},\ and\ \citenamefont {Gisin}}]{amari_towards_2010}%
  \BibitemOpen
  \bibfield  {author} {\bibinfo {author} {\bibfnamefont {A.}~\bibnamefont
  {Amari}}, \bibinfo {author} {\bibfnamefont {A.}~\bibnamefont {Walther}},
  \bibinfo {author} {\bibfnamefont {M.}~\bibnamefont {Sabooni}}, \bibinfo
  {author} {\bibfnamefont {M.}~\bibnamefont {Huang}}, \bibinfo {author}
  {\bibfnamefont {S.}~\bibnamefont {Kröll}}, \bibinfo {author} {\bibfnamefont
  {M.}~\bibnamefont {Afzelius}}, \bibinfo {author} {\bibfnamefont
  {I.}~\bibnamefont {Usmani}}, \bibinfo {author} {\bibfnamefont
  {B.}~\bibnamefont {Lauritzen}}, \bibinfo {author} {\bibfnamefont
  {N.}~\bibnamefont {Sangouard}}, \bibinfo {author} {\bibfnamefont
  {H.}~\bibnamefont {de~Riedmatten}},\ and\ \bibinfo {author} {\bibfnamefont
  {N.}~\bibnamefont {Gisin}},\ }\bibfield  {title} {\bibinfo {title} {Towards
  an eficient atomic frequency comb quantum memory},\ }\href
  {https://doi.org/10.1016/j.jlumin.2010.01.012} {\bibfield  {journal}
  {\bibinfo  {journal} {Journal of Luminescence}\ }\textbf {\bibinfo {volume}
  {130}},\ \bibinfo {pages} {1579} (\bibinfo {year} {2010})},\ \Eprint
  {https://arxiv.org/abs/0911.2145} {0911.2145} \BibitemShut {NoStop}%
\bibitem [{\citenamefont {Fukui}\ \emph {et~al.}(2018)\citenamefont {Fukui},
  \citenamefont {Tomita}, \citenamefont {Okamoto},\ and\ \citenamefont
  {Fujii}}]{fukui_high-threshold_2018}%
  \BibitemOpen
  \bibfield  {author} {\bibinfo {author} {\bibfnamefont {K.}~\bibnamefont
  {Fukui}}, \bibinfo {author} {\bibfnamefont {A.}~\bibnamefont {Tomita}},
  \bibinfo {author} {\bibfnamefont {A.}~\bibnamefont {Okamoto}},\ and\ \bibinfo
  {author} {\bibfnamefont {K.}~\bibnamefont {Fujii}},\ }\bibfield  {title}
  {\bibinfo {title} {High-threshold fault-tolerant quantum computation with
  analog quantum error correction},\ }\href
  {https://doi.org/10.1103/PhysRevX.8.021054} {\bibfield  {journal} {\bibinfo
  {journal} {Phys. Rev. X}\ }\textbf {\bibinfo {volume} {8}},\ \bibinfo {pages}
  {021054} (\bibinfo {year} {2018})},\ \Eprint
  {https://arxiv.org/abs/1712.00294} {1712.00294} \BibitemShut {NoStop}%
\bibitem [{\citenamefont {Seshadreesan}\ \emph {et~al.}(2021)\citenamefont
  {Seshadreesan}, \citenamefont {Dhara}, \citenamefont {Patil}, \citenamefont
  {Jiang},\ and\ \citenamefont {Guha}}]{seshadreesan_coherent_2021}%
  \BibitemOpen
  \bibfield  {author} {\bibinfo {author} {\bibfnamefont {K.~P.}\ \bibnamefont
  {Seshadreesan}}, \bibinfo {author} {\bibfnamefont {P.}~\bibnamefont {Dhara}},
  \bibinfo {author} {\bibfnamefont {A.}~\bibnamefont {Patil}}, \bibinfo
  {author} {\bibfnamefont {L.}~\bibnamefont {Jiang}},\ and\ \bibinfo {author}
  {\bibfnamefont {S.}~\bibnamefont {Guha}},\ }\bibfield  {title} {\bibinfo
  {title} {Coherent manipulation of graph states composed of finite-energy
  gottesman-kitaev-preskill-encoded qubits},\ }\href
  {http://arxiv.org/abs/2105.04300} {\bibfield  {journal} {\bibinfo  {journal}
  {{arXiv}:2105.04300 [quant-ph]}\ } (\bibinfo {year} {2021})},\ \Eprint
  {https://arxiv.org/abs/2105.04300} {2105.04300} \BibitemShut {NoStop}%
\bibitem [{\citenamefont {Glancy}\ and\ \citenamefont
  {Knill}(2006)}]{PhysRevA.73.012325}%
  \BibitemOpen
  \bibfield  {author} {\bibinfo {author} {\bibfnamefont {S.}~\bibnamefont
  {Glancy}}\ and\ \bibinfo {author} {\bibfnamefont {E.}~\bibnamefont {Knill}},\
  }\bibfield  {title} {\bibinfo {title} {Error analysis for encoding a qubit in
  an oscillator},\ }\href {https://doi.org/10.1103/PhysRevA.73.012325}
  {\bibfield  {journal} {\bibinfo  {journal} {Phys. Rev. A}\ }\textbf {\bibinfo
  {volume} {73}},\ \bibinfo {pages} {012325} (\bibinfo {year}
  {2006})}\BibitemShut {NoStop}%
\bibitem [{\citenamefont {Maltese}\ \emph {et~al.}(2020)\citenamefont
  {Maltese}, \citenamefont {Amanti}, \citenamefont {Appas}, \citenamefont
  {Sinnl}, \citenamefont {Lemaître}, \citenamefont {Milman}, \citenamefont
  {Baboux},\ and\ \citenamefont {Ducci}}]{maltese_generation_2020}%
  \BibitemOpen
  \bibfield  {author} {\bibinfo {author} {\bibfnamefont {G.}~\bibnamefont
  {Maltese}}, \bibinfo {author} {\bibfnamefont {M.~I.}\ \bibnamefont {Amanti}},
  \bibinfo {author} {\bibfnamefont {F.}~\bibnamefont {Appas}}, \bibinfo
  {author} {\bibfnamefont {G.}~\bibnamefont {Sinnl}}, \bibinfo {author}
  {\bibfnamefont {A.}~\bibnamefont {Lemaître}}, \bibinfo {author}
  {\bibfnamefont {P.}~\bibnamefont {Milman}}, \bibinfo {author} {\bibfnamefont
  {F.}~\bibnamefont {Baboux}},\ and\ \bibinfo {author} {\bibfnamefont
  {S.}~\bibnamefont {Ducci}},\ }\bibfield  {title} {\bibinfo {title}
  {Generation and symmetry control of quantum frequency combs},\ }\href
  {https://doi.org/10.1038/s41534-019-0237-9} {\bibfield  {journal} {\bibinfo
  {journal} {npj Quantum Information}\ }\textbf {\bibinfo {volume} {6}},\
  \bibinfo {pages} {13} (\bibinfo {year} {2020})}\BibitemShut {NoStop}%
\bibitem [{\citenamefont {Fabre}(2022)}]{doi:10.1080/09500340.2022.2073613}%
  \BibitemOpen
  \bibfield  {author} {\bibinfo {author} {\bibfnamefont {N.}~\bibnamefont
  {Fabre}},\ }\bibfield  {title} {\bibinfo {title} {Spectral single photons
  characterization using generalized hong–ou–mandel interferometry},\
  }\href {https://doi.org/10.1080/09500340.2022.2073613} {\bibfield  {journal}
  {\bibinfo  {journal} {Journal of Modern Optics}\ }\textbf {\bibinfo {volume}
  {69}},\ \bibinfo {pages} {653} (\bibinfo {year} {2022})},\ \Eprint
  {https://arxiv.org/abs/https://doi.org/10.1080/09500340.2022.2073613}
  {https://doi.org/10.1080/09500340.2022.2073613} \BibitemShut {NoStop}%
\bibitem [{\citenamefont {Le~Jeannic}\ \emph {et~al.}(2022)\citenamefont
  {Le~Jeannic}, \citenamefont {Tiranov}, \citenamefont {Carolan}, \citenamefont
  {Ramos}, \citenamefont {Wang}, \citenamefont {Appel}, \citenamefont {Scholz},
  \citenamefont {Wieck}, \citenamefont {Ludwig}, \citenamefont {Rotenberg},
  \citenamefont {Midolo}, \citenamefont {García-Ripoll}, \citenamefont
  {Sørensen},\ and\ \citenamefont {Lodahl}}]{le_jeannic_dynamical_2022}%
  \BibitemOpen
  \bibfield  {author} {\bibinfo {author} {\bibfnamefont {H.}~\bibnamefont
  {Le~Jeannic}}, \bibinfo {author} {\bibfnamefont {A.}~\bibnamefont {Tiranov}},
  \bibinfo {author} {\bibfnamefont {J.}~\bibnamefont {Carolan}}, \bibinfo
  {author} {\bibfnamefont {T.}~\bibnamefont {Ramos}}, \bibinfo {author}
  {\bibfnamefont {Y.}~\bibnamefont {Wang}}, \bibinfo {author} {\bibfnamefont
  {M.~H.}\ \bibnamefont {Appel}}, \bibinfo {author} {\bibfnamefont
  {S.}~\bibnamefont {Scholz}}, \bibinfo {author} {\bibfnamefont {A.~D.}\
  \bibnamefont {Wieck}}, \bibinfo {author} {\bibfnamefont {A.}~\bibnamefont
  {Ludwig}}, \bibinfo {author} {\bibfnamefont {N.}~\bibnamefont {Rotenberg}},
  \bibinfo {author} {\bibfnamefont {L.}~\bibnamefont {Midolo}}, \bibinfo
  {author} {\bibfnamefont {J.~J.}\ \bibnamefont {García-Ripoll}}, \bibinfo
  {author} {\bibfnamefont {A.~S.}\ \bibnamefont {Sørensen}},\ and\ \bibinfo
  {author} {\bibfnamefont {P.}~\bibnamefont {Lodahl}},\ }\bibfield  {title}
  {\bibinfo {title} {Dynamical photon–photon interaction mediated by a
  quantum emitter},\ }\href {https://doi.org/10.1038/s41567-022-01720-x}
  {\bibfield  {journal} {\bibinfo  {journal} {Nature Physics}\ }\textbf
  {\bibinfo {volume} {18}},\ \bibinfo {pages} {1191} (\bibinfo {year}
  {2022})}\BibitemShut {NoStop}%
\bibitem [{\citenamefont {Le~Jeannic}\ \emph {et~al.}(2021)\citenamefont
  {Le~Jeannic}, \citenamefont {Ramos}, \citenamefont {Simonsen}, \citenamefont
  {Pregnolato}, \citenamefont {Liu}, \citenamefont {Schott}, \citenamefont
  {Wieck}, \citenamefont {Ludwig}, \citenamefont {Rotenberg}, \citenamefont
  {Garc\'{\i}a-Ripoll},\ and\ \citenamefont {Lodahl}}]{PhysRevLett.126.023603}%
  \BibitemOpen
  \bibfield  {author} {\bibinfo {author} {\bibfnamefont {H.}~\bibnamefont
  {Le~Jeannic}}, \bibinfo {author} {\bibfnamefont {T.}~\bibnamefont {Ramos}},
  \bibinfo {author} {\bibfnamefont {S.~F.}\ \bibnamefont {Simonsen}}, \bibinfo
  {author} {\bibfnamefont {T.}~\bibnamefont {Pregnolato}}, \bibinfo {author}
  {\bibfnamefont {Z.}~\bibnamefont {Liu}}, \bibinfo {author} {\bibfnamefont
  {R.}~\bibnamefont {Schott}}, \bibinfo {author} {\bibfnamefont {A.~D.}\
  \bibnamefont {Wieck}}, \bibinfo {author} {\bibfnamefont {A.}~\bibnamefont
  {Ludwig}}, \bibinfo {author} {\bibfnamefont {N.}~\bibnamefont {Rotenberg}},
  \bibinfo {author} {\bibfnamefont {J.~J.}\ \bibnamefont
  {Garc\'{\i}a-Ripoll}},\ and\ \bibinfo {author} {\bibfnamefont
  {P.}~\bibnamefont {Lodahl}},\ }\bibfield  {title} {\bibinfo {title}
  {Experimental reconstruction of the few-photon nonlinear scattering matrix
  from a single quantum dot in a nanophotonic waveguide},\ }\href
  {https://doi.org/10.1103/PhysRevLett.126.023603} {\bibfield  {journal}
  {\bibinfo  {journal} {Phys. Rev. Lett.}\ }\textbf {\bibinfo {volume} {126}},\
  \bibinfo {pages} {023603} (\bibinfo {year} {2021})}\BibitemShut {NoStop}%
\bibitem [{\citenamefont {Mazzotta}\ \emph {et~al.}(2016)\citenamefont
  {Mazzotta}, \citenamefont {Cialdi}, \citenamefont {Cipriani}, \citenamefont
  {Olivares},\ and\ \citenamefont {Paris}}]{PhysRevA.94.063842}%
  \BibitemOpen
  \bibfield  {author} {\bibinfo {author} {\bibfnamefont {Z.}~\bibnamefont
  {Mazzotta}}, \bibinfo {author} {\bibfnamefont {S.}~\bibnamefont {Cialdi}},
  \bibinfo {author} {\bibfnamefont {D.}~\bibnamefont {Cipriani}}, \bibinfo
  {author} {\bibfnamefont {S.}~\bibnamefont {Olivares}},\ and\ \bibinfo
  {author} {\bibfnamefont {M.~G.~A.}\ \bibnamefont {Paris}},\ }\bibfield
  {title} {\bibinfo {title} {High-order dispersion effects in two-photon
  interference},\ }\href {https://doi.org/10.1103/PhysRevA.94.063842}
  {\bibfield  {journal} {\bibinfo  {journal} {Phys. Rev. A}\ }\textbf {\bibinfo
  {volume} {94}},\ \bibinfo {pages} {063842} (\bibinfo {year}
  {2016})}\BibitemShut {NoStop}%
\bibitem [{\citenamefont {Alexander}\ \emph {et~al.}(2016)\citenamefont
  {Alexander}, \citenamefont {Wang}, \citenamefont {Sridhar}, \citenamefont
  {Chen}, \citenamefont {Pfister},\ and\ \citenamefont
  {Menicucci}}]{PhysRevA.94.032327}%
  \BibitemOpen
  \bibfield  {author} {\bibinfo {author} {\bibfnamefont {R.~N.}\ \bibnamefont
  {Alexander}}, \bibinfo {author} {\bibfnamefont {P.}~\bibnamefont {Wang}},
  \bibinfo {author} {\bibfnamefont {N.}~\bibnamefont {Sridhar}}, \bibinfo
  {author} {\bibfnamefont {M.}~\bibnamefont {Chen}}, \bibinfo {author}
  {\bibfnamefont {O.}~\bibnamefont {Pfister}},\ and\ \bibinfo {author}
  {\bibfnamefont {N.~C.}\ \bibnamefont {Menicucci}},\ }\bibfield  {title}
  {\bibinfo {title} {One-way quantum computing with arbitrarily large
  time-frequency continuous-variable cluster states from a single optical
  parametric oscillator},\ }\href {https://doi.org/10.1103/PhysRevA.94.032327}
  {\bibfield  {journal} {\bibinfo  {journal} {Phys. Rev. A}\ }\textbf {\bibinfo
  {volume} {94}},\ \bibinfo {pages} {032327} (\bibinfo {year}
  {2016})}\BibitemShut {NoStop}%
\bibitem [{\citenamefont {Lütkenhaus}\ \emph {et~al.}(1999)\citenamefont
  {Lütkenhaus}, \citenamefont {Calsamiglia},\ and\ \citenamefont
  {Suominen}}]{lutkenhaus_bell_1999}%
  \BibitemOpen
  \bibfield  {author} {\bibinfo {author} {\bibfnamefont {N.}~\bibnamefont
  {Lütkenhaus}}, \bibinfo {author} {\bibfnamefont {J.}~\bibnamefont
  {Calsamiglia}},\ and\ \bibinfo {author} {\bibfnamefont {K.-A.}\ \bibnamefont
  {Suominen}},\ }\bibfield  {title} {\bibinfo {title} {Bell measurements for
  teleportation},\ }\href {https://doi.org/10.1103/PhysRevA.59.3295} {\bibfield
   {journal} {\bibinfo  {journal} {Phys. Rev. A}\ }\textbf {\bibinfo {volume}
  {59}},\ \bibinfo {pages} {3295} (\bibinfo {year} {1999})}\BibitemShut
  {NoStop}%
\bibitem [{\citenamefont {Eaton}\ \emph {et~al.}(2023)\citenamefont {Eaton},
  \citenamefont {Hossameldin}, \citenamefont {Birrittella}, \citenamefont
  {Alsing}, \citenamefont {Gerry}, \citenamefont {Dong}, \citenamefont
  {Cuevas},\ and\ \citenamefont {Pfister}}]{eaton_resolution_2023}%
  \BibitemOpen
  \bibfield  {author} {\bibinfo {author} {\bibfnamefont {M.}~\bibnamefont
  {Eaton}}, \bibinfo {author} {\bibfnamefont {A.}~\bibnamefont {Hossameldin}},
  \bibinfo {author} {\bibfnamefont {R.~J.}\ \bibnamefont {Birrittella}},
  \bibinfo {author} {\bibfnamefont {P.~M.}\ \bibnamefont {Alsing}}, \bibinfo
  {author} {\bibfnamefont {C.~C.}\ \bibnamefont {Gerry}}, \bibinfo {author}
  {\bibfnamefont {H.}~\bibnamefont {Dong}}, \bibinfo {author} {\bibfnamefont
  {C.}~\bibnamefont {Cuevas}},\ and\ \bibinfo {author} {\bibfnamefont
  {O.}~\bibnamefont {Pfister}},\ }\bibfield  {title} {\bibinfo {title}
  {Resolution of 100 photons and quantum generation of unbiased random
  numbers},\ }\href {https://doi.org/10.1038/s41566-022-01105-9} {\bibfield
  {journal} {\bibinfo  {journal} {Nature Photonics}\ }\textbf {\bibinfo
  {volume} {17}},\ \bibinfo {pages} {106} (\bibinfo {year} {2023})}\BibitemShut
  {NoStop}%
\bibitem [{\citenamefont {Ewert}\ and\ \citenamefont {van
  Loock}(2014)}]{PhysRevLett.113.140403}%
  \BibitemOpen
  \bibfield  {author} {\bibinfo {author} {\bibfnamefont {F.}~\bibnamefont
  {Ewert}}\ and\ \bibinfo {author} {\bibfnamefont {P.}~\bibnamefont {van
  Loock}},\ }\bibfield  {title} {\bibinfo {title} {$3/4$-efficient bell
  measurement with passive linear optics and unentangled ancillae},\ }\href
  {https://doi.org/10.1103/PhysRevLett.113.140403} {\bibfield  {journal}
  {\bibinfo  {journal} {Phys. Rev. Lett.}\ }\textbf {\bibinfo {volume} {113}},\
  \bibinfo {pages} {140403} (\bibinfo {year} {2014})}\BibitemShut {NoStop}%
\bibitem [{\citenamefont {Vaidman}\ and\ \citenamefont
  {Yoran}(1999)}]{PhysRevA.59.116}%
  \BibitemOpen
  \bibfield  {author} {\bibinfo {author} {\bibfnamefont {L.}~\bibnamefont
  {Vaidman}}\ and\ \bibinfo {author} {\bibfnamefont {N.}~\bibnamefont
  {Yoran}},\ }\bibfield  {title} {\bibinfo {title} {Methods for reliable
  teleportation},\ }\href {https://doi.org/10.1103/PhysRevA.59.116} {\bibfield
  {journal} {\bibinfo  {journal} {Phys. Rev. A}\ }\textbf {\bibinfo {volume}
  {59}},\ \bibinfo {pages} {116} (\bibinfo {year} {1999})}\BibitemShut
  {NoStop}%
\end{thebibliography}%

\end{document}